\documentclass[preprint]{elsarticle}

\usepackage{geometry}
\usepackage{graphicx}
\usepackage{amsmath}
\DeclareMathOperator*{\Aop}{\mathrm{\Large A}}
\usepackage{cleveref}
\usepackage{multirow}
\usepackage{arydshln}

\usepackage{xcolor}
\usepackage{ulem}
\usepackage{soul}
\definecolor{lightblue}{rgb}{0.8,0.9,1}
\sethlcolor{lightblue}

\usepackage{amssymb}
\usepackage{amsthm}

\journal{Journal of the Mechanics and Physics of Solids}

\begin{document}

\begin{frontmatter}



\title{Constitutive parameter inference using physics-based data-driven modeling in full volume datasets of intact and torn rotator cuff tendons}


\author[UmichMech,UmichMIDAS]{Carla Nathaly Villac\'is N\'u\~nez}
\author[Auburn]{Siddhartha Srivastava}
\author[UmichMech]{Ulrich Scheven}
\author[UmichOrtho]{Asheesh Bedi}
\author[SoCal]{Krishna Garikipati}
\author[UmichMech,UmichBME,UmichMacro]{Ellen M. Arruda}

\affiliation[UmichMech]{organization={Department of Mechanical Engineering, University of Michigan},
            addressline={2350 Hayward St.}, 
            city={Ann Arbor},
            postcode={48109}, 
            state={Michigan},
            country={United States}}

\affiliation[UmichMIDAS]{organization={Michigan Institute for Data \& AI in Society, University of Michigan},
            addressline={Weiser Hall, 500 Church Street, Suite 600}, 
            city={Ann Arbor},
            postcode={48109}, 
            state={Michigan},
            country={United States}}

\affiliation[Auburn]{organization={Department of Aerospace Engineering, Auburn University},
            addressline={141 Engineering Drive}, 
            city={Auburn},
            postcode={36830}, 
            state={Alabama},
            country={United States}}
            
\affiliation[UmichOrtho]{organization = {Department of Orthopedic Surgery, University of Michigan},
                addressline = {Taubman Health Care Center, 2912, 1500 E Medical Center Dr.},
                city = {Ann Arbor},
                postcode = {48109},
                state = {Michigan},
                country = {United States}}

\affiliation[SoCal]{organization={Department of Aerospace and Mechanical Engineering, University of Southern California},
            addressline={University Park Campus}, 
            city={Los Angeles},
            postcode={90089}, 
            state={California},
            country={United States}}

\affiliation[UmichBME]{organization={Department of Biomedical Engineering, University of Michigan},
            addressline={Carl A. Gerstacker Building, 2200 Bonisteel Blvd Room 1107}, 
            city={Ann Arbor},
            postcode={48109}, 
            state={Michigan},
            country={United States}}

\affiliation[UmichMacro]{organization={Program of Macromolecular Science and Engineering, University of Michigan},
            addressline={2800 Plymouth Rd}, 
            city={Ann Arbor},
            postcode={48109}, 
            state={Michigan},
            country={United States}}

\begin{abstract}

In this work, we characterized the material properties of an animal model of the rotator cuff tendon using full volume datasets of both its intact and injured states by capturing internal strain behavior throughout the tendon. Our experimental setup, involving tension along the fiber direction, activated volumetric, tensile, and shear mechanisms due to the tendon’s complex geometry. We implemented an approach to model inference that we refer to as variational system identification (VSI) to solve the weak form of the stress equilibrium equation using these full volume displacements. Three constitutive models were used for parameter inference: a neo-Hookean model, a modified Holzapfel-Gasser-Ogden (HGO) model with higher-order terms in the first and second invariants, and a reduced polynomial model consisting of terms based on the first, second, and fiber-related invariants. Inferred parameters were further refined using an adjoint-based partial differential equation (PDE)-constrained optimization framework. Our results show that the modified HGO model captures the tendon’s deformation mechanisms with reasonable accuracy, while the neo-Hookean model fails to reproduce key internal features, particularly the shear behavior in the injured tendon. Surprisingly, the simplified polynomial model performed comparably to the modified HGO formulation using only three terms. These findings suggest that while current constitutive models do not fully replicate the complex internal mechanics of the tendon, they are capable of capturing key trends in both intact and damaged tissue, using a homogeneous modeling approach. Continued model development is needed to bridge this gap and enable clinical-grade, predictive simulations of tendon injury and repair.

\end{abstract}


\begin{highlights}
    \item A single MRI experiment activates multiple deformation mechanisms simultaneously
    \item Inference using full volume displacements captures internal strain behavior
    \item Modified Holzapfel-Gasser-Ogden and polynomial models reproduce delamination behavior
    \item Fiber direction incorporation yields physically plausible constitutive responses
\end{highlights}


\end{frontmatter}


\section{Introduction}

The rotator cuff tendons are essential for shoulder stability and mobility, enabling dynamic movement while maintaining joint integrity \citep{andrews2008athlete, engelhardt2016fea}. Due to the high mobility of this joint, these tendons are prone to injury, with partial thickness rotator cuff tendon tears being a leading cause of shoulder disorders \citep{bi2024partial, finnan2010partial, matthewson2015partial}. While partial thickness tears can cause debilitating pain and disability, a large percentage of patients remain asymptomatic \citep{reilly2006partmorefull, itoi2013partial, minagawa2013age, tankala2025asymp}. Clinical management of these tears includes both conservative and surgical options, with asymptomatic tears being preferentially treated conservatively \citep{liu2018conservative, reynolds2008conservative, rudzki2008conservative}. A large percentage of partial thickness tears have been reported to progress into full thickness tears \citep{kartus2006conversion, katthagen2018retearpartial}, ultimately propagating to adjacent tendons and causing major loss of shoulder function. Hence, accurate management of tears is crucial to achieve the best possible clinical outcome and restore quality of life. While nonoperative treatment is the first-line option, surgical repair has historically been recommended for symptomatic patients with tears that span more than 50\% of the tendon’s thickness (high-grade tears), heavily based on anecdotal evidence and the belief that delaying surgical intervention might result in poorer clinical outcomes in the long term \citep{plancher2021, weber1999threshold}. However, in the absence of a mechanistic framework to predict tear progression, optimal treatment strategies remain a matter of clinical debate.

Biomechanical testing has been used to experimentally characterize the progression of rotator cuff tendon tears. In full thickness tears, changes in strain response and strain concentrations at tear tips have been associated with tear evolution around the damaged tissue \citep{andarawispuri2009strain, matthewmiller2014strain, thunes2015fea}. In contrast, predicting tear growth in partial thickness tears has been intrinsically challenging due to the two-dimensional nature of standard strain measurement techniques such as Digital Image Correlation (DIC) or marker-to-marker deformation. In particular, partial thickness tears have primarily been studied on the bursal surface of the tendon, which is visible at any shoulder position \citep{mazzoca2008repairthreshold, frisch2014bursaltear, yang2009repairthreshold}. The articular surface of the tendon, however, is obstructed by the humeral head at small abduction angles, which has driven researchers to characterize this surface almost exclusively at high abduction angles or with the humeral head resected for better visualization \citep{reilly2003partialtear, yang2009repairthreshold}. Without the ability to measure the complete loading environment of the partially injured tendon, understanding the mechanisms of tear growth is challenging, which in turn complicates the selection of the optimal treatment. This surface-limited approach overlooks the layered, heterogeneous internal architecture of the rotator cuff, which is essential to understand its full volume mechanical behavior \citep{clark1992structure, fallon2002suprastruct}.

To address these limitations, finite element models of the rotator cuff have been developed to simulate joint positions and strain distributions in partial thickness tears. From these models, we have learned that large articular strains and stresses may develop internally during abduction angles in intact and partially injured tendons \citep{luo1998fea, quental2016fea, sano2006fea, wakabayashi2003fea}, which traditional strain measurement techniques cannot fully capture. Some of these models incorporate the anisotropic and inhomogeneous properties of the rotator cuff by considering fiber orientation and dividing the tendon into multiple regions \citep{engelhardt2014fea, thunes2015fea, engelhardt2016fea, quental2016fea, matthewMiller2019fea, spracklin2019fea, williamson2023propfea}. Although these models have significantly advanced our understanding of the loading environment in intact and injured tendons, their calibration has typically relied on one-dimensional stress-strain curves or two-/three-dimensional experimental data measured on the surface of the physically excised tendon \citep{matthewMiller2019fea, szczesny2012mechprop, williamson2023propfea}. By obtaining information from samples whose native environment has been disrupted, these computational models cannot properly validate full volume mechanics. Without validation of both global and internal mechanics, these models remain limited in their clinical applicability and their ability to guide treatment decisions.

Recent advances in displacement-encoded MRI have enabled through-thickness strain measurements, revealing how microstructural differences can be captured with constitutive macro-mechanical behavior inference \citep{estrada2020mru, luetkemeyer2021constmod, scheven2020mri}. We recently showed that full volume displacements and strains measured with these MRI-based techniques reveal regions of high shear strain in the internal volume of the injured tendon \citep{villacis2025mri}. These full volume datasets highlight the importance of using the native tendon-to-bone attachments and examining areas beyond the bursal and articular surfaces of the rotator cuff. Moreover, the material heterogeneity and geometry of the rotator cuff enables the activation of multiple mechanisms in a single quasi-uniaxial test. These information-rich experimental maps are ideal for parameter inference using full volume characterization techniques like the virtual fields method (VFM) \citep{martins2018vfm} or variational system identification (VSI) \citep{wang2019vsi, wang2020vsi, wang2021vsi, wang2021vsipde}, both of which benefit from heterogeneity in experimental datasets for parameter identifiability. By relying on local full volume displacement data and the weak form of the equilibrium equations, these methods enable detailed evaluation of internal tendon mechanics.

VSI, in particular, has been validated using synthetic data and MRI-derived experimental strain maps in incompressible materials \citep{wang2021vsi}. VSI is especially useful for deriving parsimonious representations of any form of the strain energy density function, providing a powerful framework to identify the dominant deformation mechanisms activated during testing. Combining our full volume MRI datasets and VSI could be key to transforming finite element models of the rotator cuff into a thorough, clinically relevant predictive tool. Hence, in this study, we inferred parsimonious representations of candidate constitutive models using our full volume maps as input to the VSI framework. Our goal was to leverage our information-rich datasets to infer physically interpretable parameters that represent the true global nature of the rotator cuff. This characterization was designed to understand the effect of geometrical features on displacement and strain patterns of intact and torn tendons. In addition, by having access to full volume measurements, we avoided excising the tendon from its enthesis and subsequent segmentation to quantify regional properties. This approach provides an opportunity to study rotator cuff material properties in situ and opens new pathways for characterizing tears that are otherwise difficult to assess experimentally. Ultimately, this methodology has the potential to enable individualized predictive models that guide treatment selection and improve outcomes in patients with partial-thickness rotator cuff tears.

In this study, we present a comprehensive parameter inference framework that integrates experimental MRI acquisition with inverse modeling techniques. We provide an overview of the weak form of the stress equilibrium equation and quantities of interest (Section \ref{Sect:MathFramework}), followed by a description of our full volume experimental data acquisition with MRI techniques (Section \ref{Sect:MethodsExperimentalAcq}). We also describe several activation mechanisms achieved with our experiments (Section \ref{Sect:MethodsActivationMech}). VSI with partial differential equation (PDE) optimization methods (Section \ref{Sect:MethodsVSI}) is included next, followed by our candidate constitutive models (Section \ref{Sect:MethodsFunctions}). Our model setup is detailed, where we incorporated voxel-wise fiber directionality maps generated with high-resolution images (Section \ref{Sect:MethodsModelSetup}). In addition, we use a training and validation approach (Sections \ref{Sect:MethodsPartition}, \ref{Sect:MethodsErrorMaps}, and \ref{Sect:MethodsTotalError}), which provides a full volume assessment for complex material parameter inference. Finally, our results (Section \ref{Results}) enable the comparison of inferred parameters from the VSI and PDE-constrained optimization frameworks, with corresponding full volume displacements from the forward implementation, and error maps from training and validation datasets (Section \ref{Sect:DiscConc}). Altogether, this work represents the first tendon-specific modeling framework trained and validated using full volume data, incorporating native architectural and structural features into a high-fidelity inference scheme.

\section{Mathematical framework}
\label{Sect:MathFramework}
The VSI method was developed in a previous communication \citep{wang2021vsi} to infer constitutive parameters of a candidate nonlinear hyperelastic strain energy density function, $W$. In this paper, we provide the necessary details for the inverse modeling framework, as applied to biological tissue such as the rotator cuff. 

The strain energy density function can be expressed through an additive decomposition into volumetric ($W_{\text{vol}}$), isochoric ($W_{\text{iso}}$), and anisotropic ($W_{\text{aniso}}$) contributions,

\begin{equation}
    W = W_{\text{vol}} + W_{\text{iso}} + W_{\text{aniso}}
    \label{Eqn:StrainDensityAll}
\end{equation}

each of which can be expressed as a function of the invariants of the right Cauchy-Green tensor, $\mathbf{C} = \mathbf{F}^T\mathbf{F}$, where the deformation gradient is $\mathbf{F} = \mathbf{1} + \frac{\partial\textbf{u}}{\partial\mathbf{X}}$ and depends on the reference position, $\mathbf{X}$, and the displacement field, $\mathbf{u}$. The aforementioned relevant invariants are,

\begin{align}
    I_1 &= \text{tr}~\mathbf{C} \\
    I_2 &= \frac{1}{2}\left(\left(\text{tr}~\mathbf{C}\right)^2-\text{tr}\left(\mathbf{C}^2\right)\right) \\
    I_3 & = \det \mathbf{C} = J^2, \quad \text{with } J = \det\mathbf{F} \\
    I_4&=\mathbf{a}_0\cdot\mathbf{C}\mathbf{a}_0,
\end{align}

where $\textbf{a}_0$ is the fiber direction field in the undeformed configuration. The isochoric version of the first and second invariants is

\begin{align}
    \overline{I_1} &= \text{tr}~\overline{\mathbf{C}} \\
    \overline{I_2} &= \frac{1}{2}\left(\left(\text{tr}~\overline{\mathbf{C}}\right)^2-\text{tr}\left(\overline{\mathbf{C}}^2\right)\right),
\end{align}

where $\overline{\textbf{C}} = J^{-2/3}\textbf{C}$. Then, by considering a number of deformation mechanisms in terms of these invariants as possible candidates in the strain energy density function, a simple form of equation \ref{Eqn:StrainDensityAll} with polynomial dependence on the invariants becomes,
\begin{equation}
    \begin{aligned}
        W &= W_{\text{vol}}(J) + W_{\text{iso}}(\overline{I_1}, \overline{I_2}) + W_{\text{aniso}}(I_4) \\
        &= \frac{1}{2}K\left(J-1\right)^2
        +\frac{1}{2}\mu\left(\overline{I_1}-3\right)
        +\theta_1\left(\overline{I_1}-3\right)^2
        +\theta_2\left(\overline{I_1}-3\right)^3 \\
        &\quad +\theta_3\left(\overline{I_2}-3\right)^2
        +\theta_4\left(\overline{I_2}-3\right)^3
        +\theta_5\left(I_4-1\right)^2
        +\theta_6\left(I_4-1\right)^4
    \end{aligned}
    \label{Eqn:StrainEnergyDensity}
\end{equation}

where $K$ is the bulk modulus, $\mu$ is the shear modulus, and $
\theta_1$ to $\theta_6$ are coefficients of the additional terms. The first Piola-Kirchhoff stress tensor is, 

\begin{equation}
    \textbf{P}=\frac{\partial W}{\partial \textbf{F}} = \frac{\partial W}{\partial \overline{I_1}} \frac{\partial \overline{I_1}}{\partial \textbf{F}} 
    +\frac{\partial W}{\partial \overline{I_2}} \frac{\partial \overline{I_2}}{\partial \textbf{F}}
    +\frac{\partial W}{\partial I_3} \frac{\partial I_3}{\partial \textbf{F}}
    +\frac{\partial W}{\partial I_4} \frac{\partial I_4}{\partial \textbf{F}},
    \label{Eqn:PartialStress}
\end{equation}

and grouping terms into parameter–operator products, we obtain the following representation:

\begin{equation}
  \mathbf{P} = \sum_i \theta_i \,\mathbf{P}_i,  
\end{equation}

where each operator $\mathbf{P}_i(I_j,\mathbf{F})$ encodes the contribution of a specific deformation mechanism, evaluated as
$\theta_i\mathbf{P}_i(I_j,\mathbf{F}) = \frac{\partial W}{\partial I_j} \, \frac{\partial I_j}{\partial \mathbf{F}}.$ With these definitions, equation \ref{Eqn:PartialStress} can be rewritten as follows, 

\begin{multline}
    \mathbf{P} = K\mathbf{P}_K(J,\mathbf{F})
    + \mu\mathbf{P_{\mu}}(\overline{I_1},\mathbf{F}) +\theta_1\mathbf{P_1}(\overline{I_1},\mathbf{F})
    +\theta_2\mathbf{P_2}(\overline{I_1},\mathbf{F}) \\
    +\theta_3\mathbf{P_3}(\overline{I_2},\mathbf{F})
    +\theta_4\mathbf{P_4}(\overline{I_2},\mathbf{F})
    +\theta_5\mathbf{P_5}(I_4,\mathbf{F})
    +\theta_6\mathbf{P_6}(I_4,\mathbf{F}),
    \label{Eqn:PiolaKirchhoff}
\end{multline}

Then, we can implement the above expression in the weak form of the stress-equilibrium equation, which defines the basis for the VSI method,

\begin{equation}
    \int_{\Omega} \frac{\partial \mathbf{w}}{\partial \mathbf{{X}}} :\left(K\mathbf{P}_K + \mu\mathbf{P}_{\mu} +
    \theta_1 \mathbf{P}_1 + \theta_2 \mathbf{P}_2 + ... + \theta_6 \mathbf{P}_6 \right) \text{d}V - 
    \int_{\Gamma_T} \mathbf{w} \cdot \mathbf{T} \text{d}S = \mathbf{0},
    \label{Eqn:WeakForm}
\end{equation}

where $\mathbf{w}$ is a test function corresponding to the virtual displacement field, the reference volume is represented by $\Omega$, and $\Gamma_{T}$ is the boundary on the surface of the volume over which the traction, $\mathbf{T}$ is applied. We assumed negligible contribution from body forces (e.g. gravity).

To express equation \ref{Eqn:WeakForm} in the Galerkin weak form for finite element method implementation, we define a set of three-dimensional displacement vectors, $\mathbf{d}_i \in \mathbb{R}^3$, corresponding to the points $\mathbf{X}_i$ in the reference (undeformed) configuration, where $i=1,\dots,N$ (the total number of points for which data are available). The displacement field, $\mathbf{u}(\mathbf{X}) \in V$, and test function, $\mathbf{w} (\mathbf{X}) \in V_0$, are defined over each element, where $V$ is a suitable function space (e.g., a Sobolev space $H^1(\Omega, \mathbb{R})$ for the trial functions, and $V_0 \subset V$ is the space of test functions satisfying appropriate boundary conditions). These fields are approximated using interpolation functions, $N^A(\mathbf{X})$, over each element as follows,

\begin{equation}
    \mathbf{u}(\mathbf{X}) = \sum^{n_{nel}}_{A=1} N^A(\mathbf{X})\mathbf{d}_A; \qquad \mathbf{w}(\mathbf{X})=\sum^{n_{nel}}_{A=1} N^A(\mathbf{X})\mathbf{c}_A,
    \label{Eqn:InterpFunct}
\end{equation}

where $n_{nel}$ is the number of nodes in the element and $\mathbf{c}_A$ represents arbitrary values of nodal weighting functions. For an element node $A$  the Kronecker-delta property holds: $N^A(\mathbf{X}_i) = \delta_{Ai}$ because of the mesh generation scheme that locates nodes at the data points, so we have $\mathbf{d}_A = \mathbf{d}_i$. This allows us to express equation \ref{Eqn:WeakForm} as follows,

\begin{multline}
    \Aop_e \left[
    \int_{\Omega_e} \sum_A \nabla N^A \mathbf{c}_A :\left(K\mathbf{P}_K^\text{d} + \mu\mathbf{P}_{\mu}^\text{d} +
    \theta_1 \mathbf{P}_1^\text{d} + \theta_2 \mathbf{P}_2^\text{d} + ... + \theta_6 \mathbf{P}_6^\text{d} \right) \text{d}V \right. \\ - 
    \left. \int_{\Gamma_{T_e}} \sum_A N^A \mathbf{c}_A \mathbf{T} \text{d}S \right] = \mathbf{0},
    \label{Eqn:Galerkin}
\end{multline}

where $\displaystyle \Aop_e$ indicates assembly over all elements in the mesh, and $\Omega_e$ and $\Gamma_{T_e}$ represent the sub-domain and traction boundary of each element. Notice that we have dropped the subscript $i$. The stress operators in equation \ref{Eqn:Galerkin} have been written with a superscript, $(\cdot)^\text{d}$, which indicates fields have been evaluated from the known data. The residual vector is built from the above equation, after considering the degrees of freedom of the test function and the finite element method definitions,

\begin{equation}
    \mathbf{R}=-\Aop_e \left[
    \int_{\Omega_e} \nabla \mathbf{N} \cdot \left(K\mathbf{P}_K^\text{d} + \mu\mathbf{P}_{\mu}^\text{d} +
    \theta_1 \mathbf{P}_1^\text{d} + \theta_2 \mathbf{P}_2^\text{d} + ... + \theta_6 \mathbf{P}_6^\text{d} \right) \text{d}V - 
    \int_{\Gamma_{T_e}} \mathbf{N} \otimes \mathbf{T} \text{d}S \right],
    \label{Eqn:Residual}
\end{equation}

where $\mathbf{N} = \{N^1, ..., N^{n_{nel}}\}$ represents the vector of interpolation functions for each element.

\section{Methods}
\label{Methods}
\subsection{Full volume experimental data acquisition}
\label{Sect:MethodsExperimentalAcq}

\subsubsection{Sample preparation}
\label{Sect:MethodsSamplePrep}

Our experiments were performed in ovine infraspinatus tendons, which have been validated as an animal proxy for human supraspinatus tendons \citep{coleman2003animals, lebaschi2016animals, schneeberger2002animals}. An exhaustive sample preparation, fixture fitting and loading protocols have been described in our previous publication \citep{villacis2025mri} and briefly summarized in this paper. Fresh sheep shoulders ($n=12$) were procured from a local butcher and dissected to isolate the infraspinatus tendon, preserving the humeral attachment and intramuscular tendon section. Specimens were fit into custom-made 3D printed fixtures at a quasi neutral position of $\sim$0 degrees abduction and rotation, with fascicles aligned in the direction of loading. Each sample-fixture assembly was mounted into an MRI compatible device, in which specimens underwent cyclic displacement in the longitudinal direction at 0.33 Hz. Two clinically relevant conditions were evaluated, intact state and torn state, with a 75\% bursal footprint detachment (type III tear). Each sample condition underwent two physiologically relevant stretches of 1 mm and 2 mm, which represented submaximal and supramaximal human shoulder force ranges, respectively. In total, four consecutive tests - intact, 1 mm and 2 mm, followed by torn, 1 mm and 2 mm - were performed. Viscoelastic effects were minimized by a pre-conditioning step which preceded each test. Figures and details of the experimental device can be found in our previous communications \citep{luetkemeyer2021constmod, scheven2020mri, villacis2025mri}.

\subsubsection{MRI sequence for strain acquisition }
\label{Sect:MethodsMRISequence}

The loading protocol was synchronized with a previously validated custom displacement encoding MRI sequence, termed APGSTEi \citep{scheven2020mri}. Briefly, during MRI acquisition with this protocol, complex-valued MRI images of the sample were generated. These images contained information of each voxel phase, modulo 2$\pi$, which was inversely proportional to the user-defined encoding wavelength, $\lambda$, and directly proportional to the voxel displacement measured from undeformed to stretched states. Displacements could then be mapped from unwrapping the phase image. A low resolution grid of 96x32x32 voxels, with a 0.39 mm x 0.65 mm x 1 mm voxel size, was used for our displacement encoded image acquisition. We used wavelengths of 1 mm and 3 mm for tendon elongations of 1 mm and 2 mm, respectively. Before each loading protocol, a high-resolution anatomical image dataset was acquired with a 3D gradient echo protocol, using a grid of 192x128x128 voxels sized 0.2917 mm x 0.1875 mm x 0.25 mm.

\subsubsection{Data processing pipeline}
\label{Sect:MethodsDataProc}

\begin{figure}[h]
    \centering
    \includegraphics[width=\linewidth]{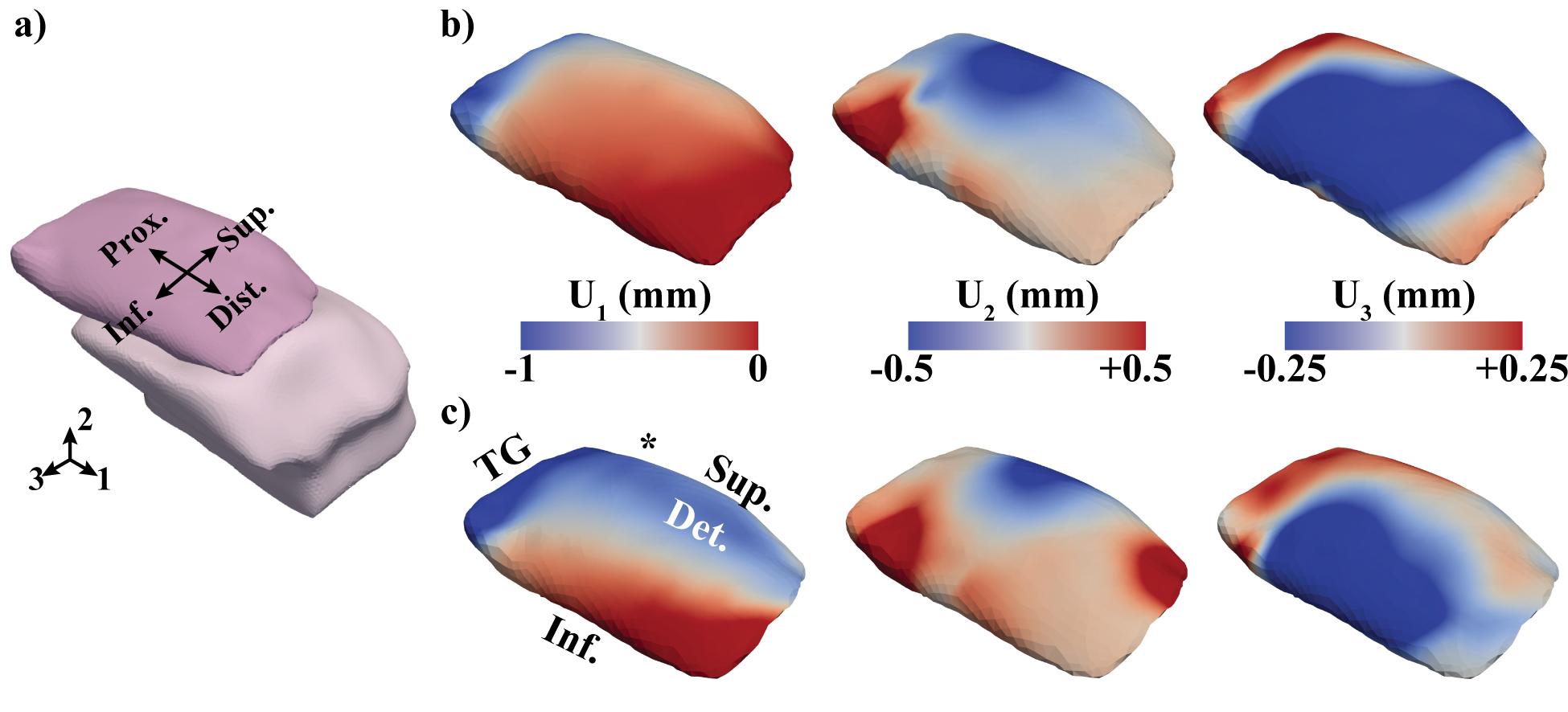}
    \caption{Representative full volume displacement maps obtained with MRI-based strain acquisition protocol. a) Three-dimensional orientation of a sample (right shoulder), showing longitudinal (1), thickness (2), and transverse (3) directions, and anatomical axes. b) Intact and c) torn displacement maps in the longitudinal (left column), thickness (middle column), and transverse (right column) directions. *TG = tendon grip, Sup. = Superior, Inf. = Inferior, Det. = Detached tissue band.}
    \label{fig:RepresentativeDisps}
\end{figure}

Full volume complex valued phase maps were unwrapped into displacement maps using an in-house developed data processing pipeline written in MATLAB (The MathWorks, Inc., version 2023b). As part of this procedure, we created binary masks which isolated the intact and torn tendon regions. These masks were converted into smooth three-dimensional meshes using Meshmixer (Autodesk, Inc., version 3.5) and Hypermesh (Altair Engineering Inc., version 2019). Full volume displacement and strain values were interpolated into each mesh with a custom Python (The Python Software Foundation, version 3.11.0) algorithm. A representative example of intact and torn full volume displacements acquired with our experimental protocol is shown in Figure \ref{fig:RepresentativeDisps}-b and c, respectively. Anatomical axes of a tendon (right shoulder) are defined in Figure \ref{fig:RepresentativeDisps}-a for future reference. Longitudinal, thickness, and transverse directions are used in this study and are oriented along the 1, 2, and 3 axis, respectively.

\subsection{Activation mechanisms and features of interest}
\label{Sect:MethodsActivationMech}

\begin{figure}[h!]
    \centering
    \includegraphics[width=\linewidth]{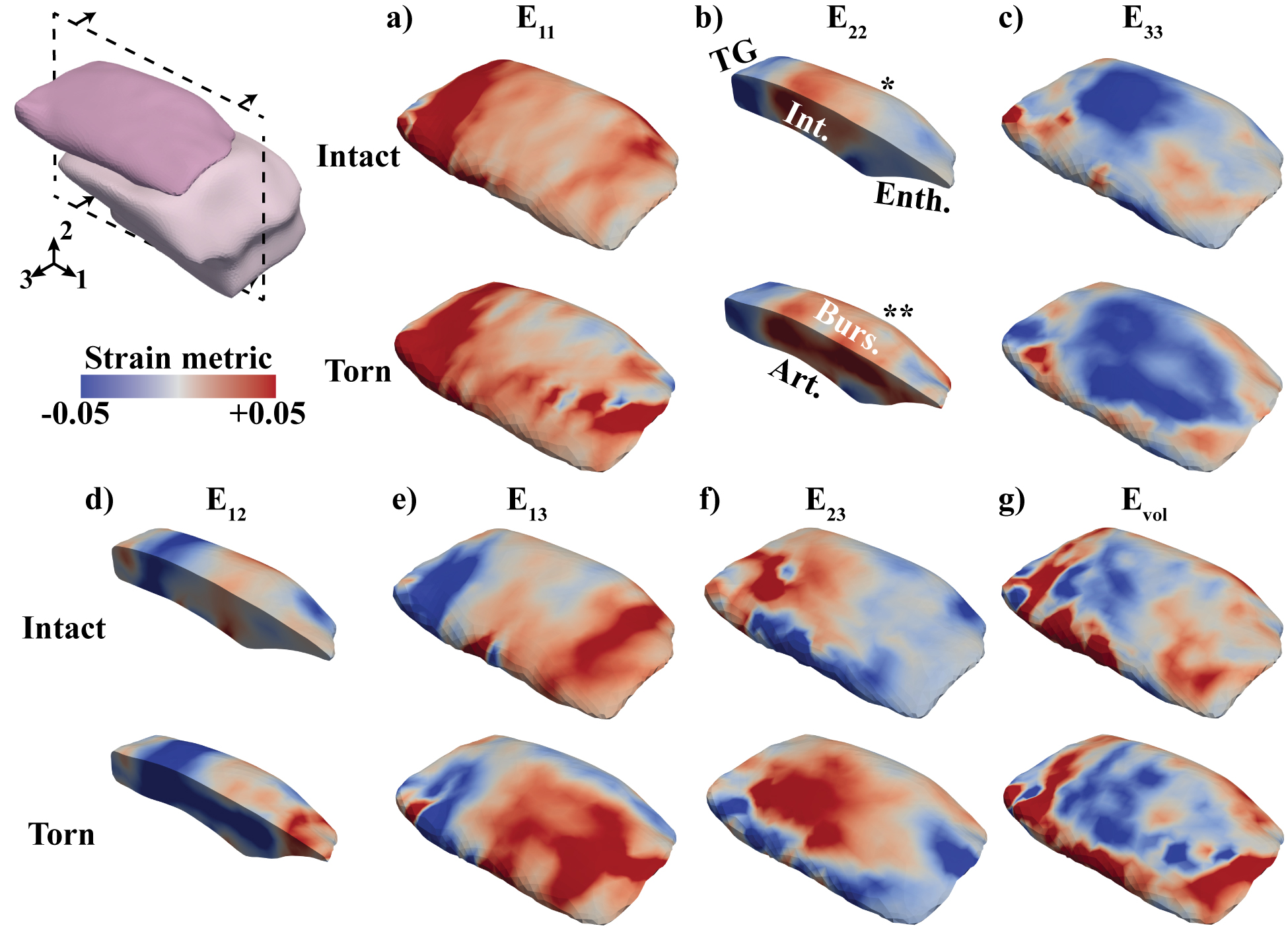}
    \caption{Mechanisms activated with our quasi-uniaxial tensile experimental setup. Normal strain components shown in a), b), and c), and shear strain components shown in d), e), and f) of the Lagrangian strain tensor are activated both in the intact and torn states. The torn condition activates a pronounced shear strain response. The volumetric strain component ($\mathrm{E_{vol}}=J-1$) in g) indicates that the material has a degree of compressibility. The dashed lines in the three-dimensional representation at the top left corner represent the sectioning plane for frames b) and d). *TG = tendon grip, Int. = internal, Enth. = enthesis. **Art. = articular, Burs. = bursal.}
    \label{fig:ActivationMechanisms}
\end{figure}

Our experimental setup, which applied tension in the direction of the fibers, activated all normal and shear-related components of the Lagrangian strain tensor in all tendons (Figure \ref{fig:ActivationMechanisms}). Positive longitudinal strain was prominent near the tendon gripper, as expected for the applied loading, but the other normal strain components were also activated. In addition, torn tendons showed distinct features corresponding to the 1-2 and 1-3 shear strain components. This multi-axial activation pattern with a single loading direction reflects the complex physiological loading environment of tendons, where tensile and shear strains coexist during joint motion, particularly within injured tendons.

High shear strain bands were observed in the internal regions of the torn tendon (Figure \ref{fig:ActivationMechanisms}-d) and the attached/detached tissue boundaries (Figure \ref{fig:ActivationMechanisms}-e), a phenomenon not previously registered or considered a critical feature in tendon tear growth. As reported in our prior work \citep{villacis2025mri}, these regions of high shear strain concentration indicate a delamination failure mechanism characterized by the separation of tendon layers at the cut-uncut interface. Specifically, the steep gradients in displacement between the attached and detached adjacent regions of the tendon (Figure \ref{fig:RepresentativeDisps}-c, first column) create these localized strain concentrations that promote crack propagation via delamination. Thus, resolving both high shear concentrations and interfacial displacement gradients was essential in our computational model to faithfully reproduce the physiologic failure mechanisms observed in tendon injury.

\subsection{Parameter inference using variational system identification}
\label{Sect:MethodsVSI}
The VSI method relies on the weak form of the stress equilibrium equation to relate basis operators and deformation mechanisms. From equation \ref{Eqn:Residual}, the target value of our model would be the known traction term assembled at the boundary, which we define as $\mathbf{y}$,

\begin{equation}
    \mathbf{y} = \Aop_e \int_{\Gamma_{T_e}} \mathbf{N} \otimes \mathbf{T} \text{d}S
    \label{Eqn:TargetValue}
\end{equation}

We also define matrix operators, $\Xi_\alpha$, from the contributions of each one of the candidate terms in the strain energy density function, expressed in equation \ref{Eqn:Residual} as the different contributions from the first Piola-Kirchhoff stress tensor,

\begin{equation}
    \Xi_{\alpha} = \Aop_e \int_{\Omega_e} \nabla \mathbf{N} \cdot \mathbf{P}_{\alpha}^d \text{d}V, \quad \alpha = K, \mu, 1, 2, ..., 6
    \label{Eqn:BasisOperators}
\end{equation}

Replacing equations \ref{Eqn:TargetValue} and \ref{Eqn:BasisOperators} in equation \ref{Eqn:Residual}, we have the residual in vector multiplication form,

\begin{equation}
    \mathbf{R}(K,\mu,\theta) = \mathbf{y} - \left[ \Xi_K \quad \Xi_{\mu} \quad \Xi_1 \quad \Xi_2 \quad \Xi_3 \quad \Xi_4 \quad \Xi_5 \quad \Xi_6\right] 
    \left\{
        \begin{aligned}
        & K \\
        & \mu \\
        & \theta_1 \\
        & \theta_2 \\
        & \theta_3 \\
        & \theta_4 \\
        & \theta_5 \\ 
        & \theta_6
    \end{aligned}
    \right\},
    \label{Eqn:ResidualMatForm}
\end{equation}

where $\theta = \langle \theta_1, ..., \theta_6\rangle^T$, and we define the complete matrix of basis operators and coefficient vector as,

\begin{align}
    \Xi &= \left[ \Xi_K \quad \Xi_{\mu} \quad \Xi_1 \quad \Xi_2 \quad \Xi_3 \quad \Xi_4 \quad \Xi_5 \quad \Xi_6\right] \\
    \mathbf{c} &= \left[K \quad \mu \quad \theta_1 \quad \theta_2 \quad \theta_3 \quad \theta_4 \quad \theta_5 \quad \theta_6\right]^T = [K \quad \mu \quad \theta]^T
\end{align}

The VSI method infers the governing physics of the material by identifying a sparse set of nonzero coefficients from the candidate strain energy terms. This is achieved by minimizing a loss function, $\ell_\text{d}$,

\begin{equation}
    \ell_\text{d}(K,\mu,\theta)=||\mathbf{R}(K,\mu,\theta)||^2,
    \label{Eqn:LossFromData}
\end{equation}

which is equivalent to:

\begin{equation}
    \ell_\text{d}(\mathbf{c})=||\mathbf{y}-\Xi\mathbf{c}||^2,
    \label{Eqn:LossFromDataC}
\end{equation}

Given the linear dependence of the residual on the coefficients, this minimization problem can be solved using standard linear regression,

\begin{equation}
    \mathbf{c} = \arg \min_{\tilde{\mathbf{c}}} \left( || \mathbf{y} - \Xi \tilde{\mathbf{c}} ||^2 \right)
\end{equation}

Then, 

\begin{equation}
    \{K, \mu, \theta\} = \arg \min_{\tilde{K}, \tilde{\mu}, \tilde{\theta}} \left( \ell_\text{d} \left(\tilde{K}, \tilde{\mu}, \tilde{\theta} \right) \right)
\end{equation}

Assuming that $\Xi$ has full rank (which is typical in an overdetermined system), the least-squares solution is given by the Moore–Penrose pseudoinverse,

\begin{equation}
    \left\{
    \begin{aligned}
        K & \\
        \mu & \\
        \theta &
    \end{aligned}
    \right\}
    = \left(\Xi^{T}\Xi\right)^{-1}\Xi^{T} \mathbf{y}
\end{equation}

\subsubsection{Addressing ill-defined system}

The ill-posedness of the inverse problem arises from the manner in which boundary conditions are enforced. Specifically, Dirichlet conditions are prescribed directly from the experimentally measured displacement field. Owing to the elliptic nature of the governing equations, a uniform rescaling of all material parameters by a positive scalar leaves the solution field unchanged. In such cases, the residual remains identically zero because no boundary traction term is present to break this invariance.

To resolve this indeterminacy, we introduce an additional loss term that enforces traction balance in a weak sense. Since the target values in equation \ref{Eqn:WeakForm} are expressed in weak form, the system as posed admits infinitely many optimal solutions. By augmenting the formulation with a force-balance constraint, derived from experimental boundary traction measurements, we regularize the problem and render it well-posed. The resulting total loss function is defined as

\begin{equation}
    \ell_T = \ell_\text{d} + \beta\ell_F,
\end{equation}
where $\ell_\text{d}$ is the displacement-related loss from equation \ref{Eqn:LossFromData}, $\ell_F$ is the force-related loss, and $\beta$ is a coefficient to equilibrate different orders of magnitude from the two loss functions, chosen at the discretion of the researchers. $\ell_F$ can be defined as,

\begin{equation}
    \ell_F = \left(F_{\text{ext}} - \int_{\Omega_T} \mathbf{T} \cdot \mathbf{n} \text{d}S\right)^2,
\end{equation}

where $F_{ext}$ is the known external load at the boundary measured from the experiment. 

To ensure both $\ell_\text{d}$ and $\ell_F$ are dimensionless and can be correctly included into a single loss function, we use the known external load,

\begin{equation}
    \ell_T = \frac{1}{F_{\text{ext}}^2} \left[\frac{1}{N_{\text{dofs}}}\left|\mathbf{R}(K,\mu,\theta)\right|^2+\beta\left( F_{\text{ext}} - \int_{\Omega_T} \mathbf{T} \cdot \mathbf{n} \text{d}S \right)^2\right],
    \label{Eqn:TotalLossVSI}
\end{equation}
 
where $N_{\text{dofs}}$ is the number of degrees of freedom in the system, and was included as a coefficient of $\ell_\text{d}$ to account for the influence of the mesh size. For multiple loading cases, parameter inference is expressed in the form

\begin{multline}
     \{ K, \mu, \theta\} = \arg \min_{\tilde{K}, \tilde{\mu}, \tilde{\theta}} \sum_{k=1}^M \frac{1}{(F_{\text{ext}})_k^2} \left[\frac{1}{(N_{\text{dofs}})_k}\left|\mathbf{R}_k(K,\mu,\theta)\right|^2\right. \\
     \left. + \beta\left( (F_{\text{ext}})_k - \int_{\Omega_T} \mathbf{T}_k \cdot \mathbf{n}_k \text{d}S \right)^2 \right],
    \label{Eqn:MinimizationVSI}
\end{multline}

where $M$ represents the number of loading cases.

\subsubsection{Parameter suppression with step-wise regression}

The VSI method employs an iterative sparse regression strategy in which inactive deformation mechanisms are systematically eliminated by setting their corresponding coefficients in the set $\{K, \mu, \theta\}$ to zero. Following earlier work \citep{wang2021vsipde}, we adopt a stepwise regression procedure combined with a statistical F-test to evaluate the significance of coefficients at each iteration. This approach begins with an initially large library of strain energy density function and progressively prunes terms whose removal produces negligible change in the loss, thereby converging to a parsimonious representation.

The F-test quantifies the normalized percent change in the loss between successive iterations,

\begin{equation}
    F = \frac{\frac{\ell_T^{i+1}-\ell_T^i}{p^i-p^{i+1}}}{\frac{\ell_T^i}{m-p^i}},
\end{equation}

where $p^i$ accounts for the number of operators at iteration $i$, and $m$ is the number of parameters in the initial strain energy density function. Reduction in parameter count directly corresponds to a decrease in model complexity. The iterative elimination process is terminated once the computed F-value exceeds a predefined tolerance, ensuring that further removal of terms would incur a substantial increase in loss.

\subsubsection{Parameter optimization with partial differential equation (PDE)-constrained solver}

In highly incompressible materials such as the rotator cuff tendons, correctly inferring the bulk modulus, $K$, can be a challenging task due to the vanishing nature of the $(J-1)$ term in the strain energy density function. To ensure the best possible fit to our experimental datasets, a PDE-constrained minimization step was implemented following VSI inference and identification of activation mechanisms. In this step, we define a loss function corresponding to the known displacement field, $\mathbf{u}$,

\begin{equation}
    \ell_u = \int_{\Omega}\left( \mathbf{u}^{\text{FE}} - \mathbf{u} \right)^2 \text{d}v
\end{equation}

where $\mathbf{u}_{\text{FE}}$ is the forward solution to the finite element problem with the current parameters. Parameter refinement is carried out by error minimization on this loss function,

\begin{equation}
    \{ K,\mu,\theta \} = \arg \min_{\hat{K}, \hat{\mu}, \hat{\theta}} \sum_{k=1}^M \int_{\Omega}\left|\left| \mathbf{u}_k^{\text{FE}} - \mathbf{u}_k \right|\right|^2 \text{d}v, \qquad \text{subject to: } \mathbf{R}(\mathbf{u}_k^{\text{FE}};\hat{K}, \hat{\mu}, \hat{\theta}) = \mathbf{0},
\end{equation}

where M represents the total number of loading cases, $\mathbf{u}_k^{\text{FE}}$ corresponds to the displacement field for the $k$th load step, obtained with forward solution to the finite element problem using the parameters $\{\hat{K}, \hat{\mu}, \hat{\theta}\}$, and $\mathbf{u}_k$ is the known experimental displacement field. We notice that this system suffers the same indeterminacy of linear coefficients up to a scaling factor under Dirichlet boundary conditions as does model inference by VSI. This indeterminacy also can be mollified by inclusion of a force balance term in the total PDE-constrained loss, $\ell_{opt}$, 

\begin{equation}
    \ell_{opt} = \ell_u + \beta\ell_F,
\end{equation}

which leads to parameter refinement by minimization in its dimensionless form,

\begin{equation}
    \{ K,\mu,\theta \} = \arg \min_{\hat{K}, \hat{\mu}, \hat{\theta}} \sum_{k=1}^M \left[ \frac{1}{V_k}\int_\Omega\frac{ \left|\left| \mathbf{u}_k^{\text{FE}} - \mathbf{u}_k \right|\right|^2 }{||(\mathbf{u}_{\text{max}})_k||^2}\text{d}v + \beta \left(1-\frac{1}{(F_{\text{ext}})_k} \int_{\Gamma_T} \mathbf{T}_k\cdot \mathbf{n}_k \text{d}S \right)^2 \right],
\end{equation}

where $|(\mathbf{u}_{\text{max}})_k|$ is the magnitude of the maximum displacement in each set, and $V_k$ is the volume of each mesh. This PDE-constrained minimization requires a gradient of the total loss function, $\ell_{opt}$, with respect to the inferred parameters. This gradient can be obtained by the adjoint approach, which needs a single solve of the linear PDE-constrained adjoint equation and a nonlinear solution for the set $\{ \hat{K}, \hat{\mu}, \hat{\theta} \}$.

It is important to notice that the implementation of the VSI and PDE-constrained optimization algorithms into the full volume datasets was carried out entirely in Python, using the FEniCS, Dolfin, and UFL libraries \citep{logg2010dolfin, logg2012automated, alnaes2015fenics}. For the PDE-constrained optimization, in particular, we chose the dolfin-adjoint software library \citep{mitusch2019dolfin}, selecting the L-BFGS-B optimization algorithm from the SciPy package \citep{virtanen2020scipy}. 

\subsection{Candidate constitutive models}
\label{Sect:MethodsFunctions}
Several strain energy density functions have been employed to capture the mechanics of rotator cuff tendons, including fiber-reinforced formulations with Mooney–Rivlin, neo-Hookean, or exponential-type matrix components \citep{engelhardt2016fea, ferrer2020model, matthewMiller2019fea, thunes2015fea}, as well as more advanced anisotropic models that incorporate fiber dispersion effects \citep{garcia2024fea, quental2016fea, williamson2023propfea}. Other tissues, such as the anterior cruciate ligament and the medial collateral ligament, have been modeled with an anisotropic extension of the eight-chain model for rubber elasticity \citep{arruda1993rubber, luetkemeyer2021constmod, marchi2018medial}. In this study, we selected three models of the strain energy density function to perform parameter inference: neo-Hookean (NH), modified Holzapfel–Gasser–Ogden (m-HGO), and polynomial (Pol). The NH formulation was included to assess how well a simplified constitutive model, commonly used to describe the nonlinear behavior of polymeric materials -- could capture complex tendon mechanics, serving as a baseline model with the minimal number of terms needed to generate a nonlinear response, as follows,

\begin{equation}
    W_{\text{NH}} = \frac{1}{2}K\left( J-1 \right)^2 + \frac{1}{2}\mu\left( \overline{I_1}-3 \right),
    \label{Eqn:NeoHookean}
\end{equation}

which yields the following form of the first Piola-Kirchhoff stress tensor,

\begin{equation}
    \mathbf{P}_\text{NH} = KJ\left( J-1 \right)\mathbf{F}^{-T} + \mu J^{-\frac{2}{3}} \left( \mathbf{F} - \frac{1}{3} I_1 \mathbf{F}^{-T}\right)
    \label{Eqn:NHStress}
\end{equation}

The incompressible version of the Holzapfel-Gasser-Ogden (HGO) model has been previously employed to model biological tissues \citep{holzapfel2000hgomodel}, including the rotator cuff tendons. While the latest versions of the three-dimensional finite element models of the rotator cuff have implemented this function in its incompressible form \citep{garcia2024fea, quental2016fea, williamson2023propfea}, we have introduced compressibility in the anisotropic portion, as described in the work by \citep{nolan2014hgomodel}, for the correct computational implementation of the finite element formulation in Python. We have further added high-order terms which depend on the first and second invariants to evaluate the activation of different mechanisms. The final version of the m-HGO model is,

\begin{equation}
    \begin{aligned}
        W_{\text{m-HGO}} &= \frac{1}{2}K\left( J-1 \right)^2
        +\frac{1}{2}\mu\left( \overline{I_1}-3 \right) \\
        &+ \frac{k_1}{2k_2} \left( \exp \left[ k_2\left( \kappa I_1 + \left( 1-3\kappa \right)I_4 - 1 \right)^2 \right] - 1\right) \\
        & + \theta_1 \left( \overline{I_1} - 3\right)^2
        + \theta_2 \left( \overline{I_1} - 3\right)^3
        + \theta_3 \left( \overline{I_1} - 3\right)^4 \\
        & + \theta_4 \left( \overline{I_2} - 3\right)^2
        + \theta_5 \left( \overline{I_2} - 3\right)^3
        + \theta_6 \left( \overline{I_2} - 3\right)^4, 
    \end{aligned}
    \label{Eqn:HGOEnergy}
\end{equation}

where $k_1$ corresponds to the stiffness of the collagen fiber phase, $k_2$ represents the fiber nonlinear response, $\kappa$ measures the degree of fiber dispersion and ranges from 0 (perfect alignment) to 1/3 (isotropically distributed or no preferred fiber alignment), and $\theta_1$ to $\theta_3$ and $\theta_4$ to $\theta_6$ are coefficients which depend on the first and second invariants, respectively. The corresponding expression for the first Piola-Kirchhoff stress is,

\begin{equation}
    \begin{aligned}
        \mathbf{P}_{\text{m-HGO}} &= KJ\left( J-1 \right)\mathbf{F}^{-T} 
        + \mu J^{-\frac{2}{3}} \left( \mathbf{F} - \frac{1}{3} I_1 \mathbf{F}^{-T}\right) \\
        &+ 2k_1I_{14} \exp \left( k_2 I_{14}^2 \right) \left[ \kappa \mathbf{F} + \left( 1-3\kappa \right)\mathbf{F}\left( \mathbf{a_0} \otimes \mathbf{a_0} \right) \right] \\ 
        &+ 2J^{-\frac{2}{3}}\left[2\theta_1\left( \overline{I_1} -3 \right) + 3\theta_2\left( \overline{I_1} -3 \right)^2 + 4\theta_3\left( \overline{I_1} -3 \right)^3\right] \left( \mathbf{F} - \frac{1}{3} I_1 \mathbf{F}^{-T}\right) \\
        &+ 2J^{-\frac{4}{3}} 
        \left[2\theta_4\left( \overline{I_2} -3 \right) + 3\theta_5\left( \overline{I_2} -3 \right)^2 + 4\theta_6\left( \overline{I_2} -3 \right)^3\right] \left( I_1\mathbf{F} - \mathbf{FC} -\frac{2}{3} I_2 \mathbf{F}^{-T}\right),
    \end{aligned}
    \label{Eqn:HGOStress}
\end{equation}

where we have included the modified fiber invariant, $I_{14} = \kappa I_1 + \left(1-3\kappa\right)I_4 -1$. Finally, we assembled a polynomial model with 11 candidate terms, which VSI is able to evaluate. Compressibility was assigned to the fourth invariant, to keep an analogous representation to the m-HGO model, as follows,

\begin{multline}
    W_{\text{Pol}} = \frac{1}{2}K\left( J-1 \right)^2
    + \frac{1}{2}\mu\left( \overline{I_1}-3 \right) 
    + \theta_1 \left( \overline{I_1} - 3\right)^2
    + \theta_2 \left( \overline{I_1} - 3\right)^3
    + \theta_3 \left( \overline{I_1} - 3\right)^4 \\
    + \theta_4 \left( \overline{I_2} - 3\right)^2
    + \theta_5 \left( \overline{I_2} - 3\right)^3
    + \theta_6 \left( \overline{I_2} - 3\right)^4 
    + \theta_7 \left( I_4 - 1 \right)^2
    + \theta_8 \left( I_4 - 1 \right)^4
    + \theta_9 \left( I_4 - 1 \right)^6
    \label{Eqn:PolEnergy}
\end{multline}

The corresponding first Piola-Kirchhoff stress is written as,

\begin{equation}
    \begin{aligned}
        \mathbf{P}_{\text{Pol}} &= KJ\left( J-1 \right)\mathbf{F}^{-T}
        + \mu J^{-\frac{2}{3}} \left( \mathbf{F} - \frac{1}{3} I_1 \mathbf{F}^{-T}\right) \\
        &+ 2J^{-\frac{2}{3}}\left[2\theta_1\left( \overline{I_1} -3 \right) + 3\theta_2\left( \overline{I_1} -3 \right)^2 + 4\theta_3\left( \overline{I_1} -3 \right)^3\right] \left( \mathbf{F} - \frac{1}{3} I_1 \mathbf{F}^{-T}\right) \\
        &+ 2J^{-\frac{4}{3}} 
        \left[2\theta_4\left( \overline{I_2} -3 \right) + 3\theta_5\left( \overline{I_2} -3 \right)^2 + 4\theta_6\left( \overline{I_2} -3 \right)^3\right] \left( I_1\mathbf{F} - \mathbf{FC} -\frac{2}{3} I_2 \mathbf{F}^{-T}\right) \\
        &+ 4\left[\theta_7\left( I_4 - 1 \right) + 2\theta_8\left( I_4 - 1 \right)^3 + 3\theta_9\left( I_4 - 1 \right)^5\right] \mathbf{F} \left( \mathbf{a_0} \otimes \mathbf{a_0} \right)
    \end{aligned}
    \label{Eqn:PolStress}
\end{equation}

\subsection{Model setup and evaluation} 
\label{Sect:MethodsModelSetup}

VSI with parameter suppression via step-wise regression was applied to our full volume datasets to infer 12 tendon-specific models, for each constitutive function, followed by parameter refinement with PDE-constrained optimization. We describe the boundary conditions, generation of a fiber direction map, mesh convergence, and training and validation procedures below.

\subsubsection{Boundary conditions}
\label{Sect:MethodsBCs}

Dirichlet boundary conditions which matched the experimental data were applied to three areas of the tendon: the enthesis or tendon-to-bone attachment, the area of contact between the tendon and the humeral head, and the area of application of load at the tendon grip. Rather than explicitly simulating contact, which can introduce numerical challenges, we leveraged the available experimental measurements to prescribe boundary conditions directly.

\begin{figure}[h!]
    \centering
    \includegraphics[width=\linewidth]{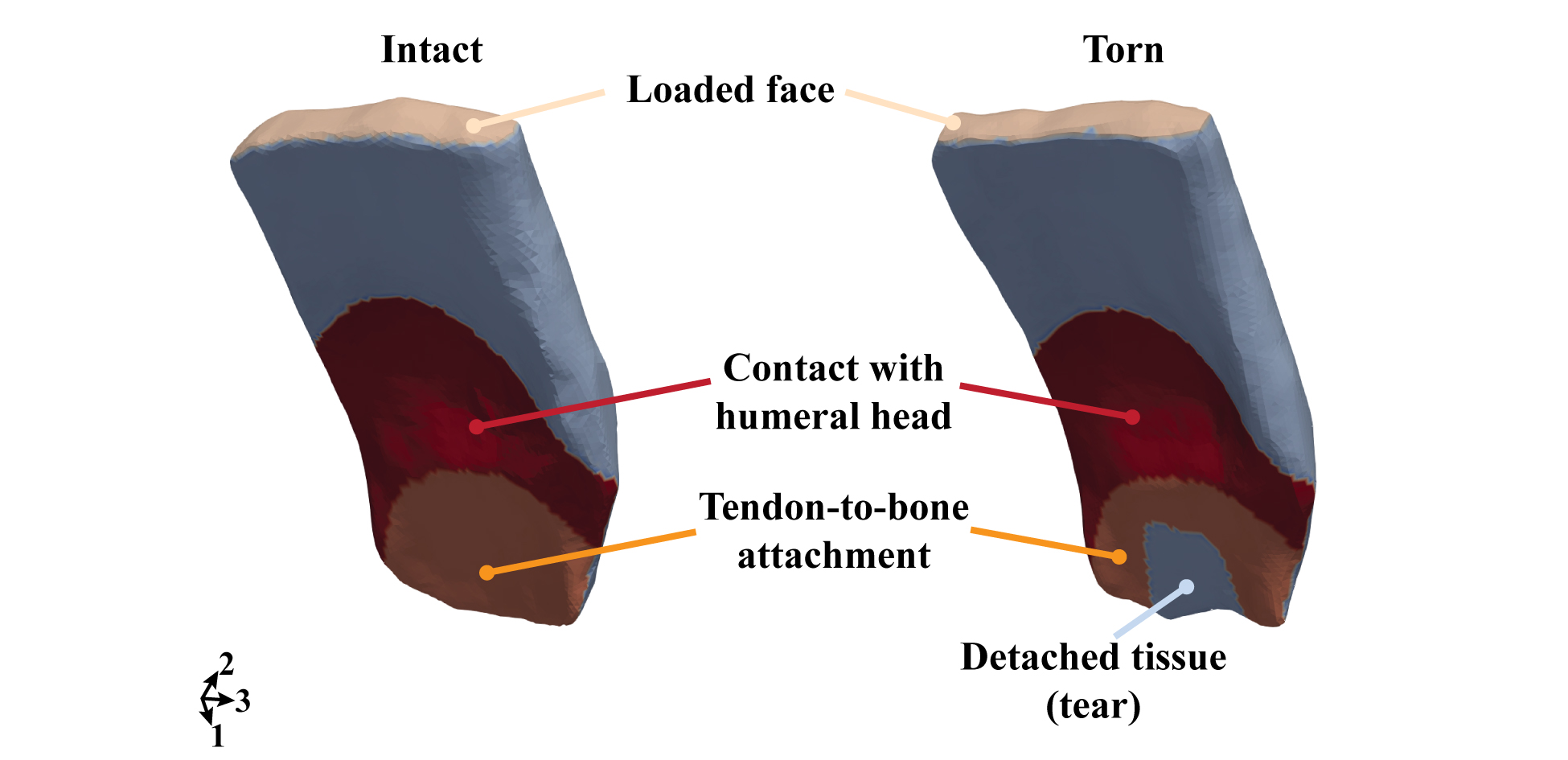}
    \caption{Boundary conditions applied to the intact and torn states of a representative sample. Equations of the humeral head and enthesis (tendon-to-bone attachment) boundaries shown in dark red and dark orange were extracted from high-resolution images, whereas the tear region shown in light blue was found with Paraview.}
    \label{fig:BCs}
\end{figure}

The face where load was applied (Figure \ref{fig:BCs}, pale region) included all surface elements close to the tendon grip, whose normals were approximately aligned to the negative 1 direction (axis shown in Figures \ref{fig:BCs} and \ref{fig:FiberDirections}). The tendon-to-bone attachment region and area of contact between the tendon and the humeral head were identified using high-resolution anatomical images. For each slice in the 1-2 plane (Figure \ref{fig:FiberDirections}-a), two points were selected, one which defined the most proximal boundary of the enthesis, and the other which defined the humeral head end. Two three-dimensional point clouds were obtained from all slices, which corresponded to the boundaries of the enthesis and the humeral head end, respectively. A polynomial function in the 1-3 plane was fit to each point cloud using the \textit{polyfit} function to obtain a smooth analytical description of both boundaries (Figure \ref{fig:BCs}, dark red and dark orange regions). Because both the intact and torn loading cases were used for parameter inference, the fitting was performed separately for the intact and torn geometries. Each tendon therefore had two boundaries (enthesis and humeral head end) × two loading conditions (intact and torn), resulting in four polynomial boundary equations. In the torn condition, tear boundaries were also selected from visualization in Paraview (Sandia National Laboratories, Kitware Inc, Los Alamos National Laboratory, version 5.12.1), and fit to an additional polynomial curve. These tear curves were subtracted from the torn entheses. Visual verification of the selected boundary elements was performed in Paraview. A representative tendon with its corresponding boundary conditions is depicted in Figure \ref{fig:BCs}.

\subsubsection{Voxel-wise fiber direction implementation}
\label{Sect:MethodsFiber}

\begin{figure}[h!]
    \centering
    \includegraphics[width=\linewidth]{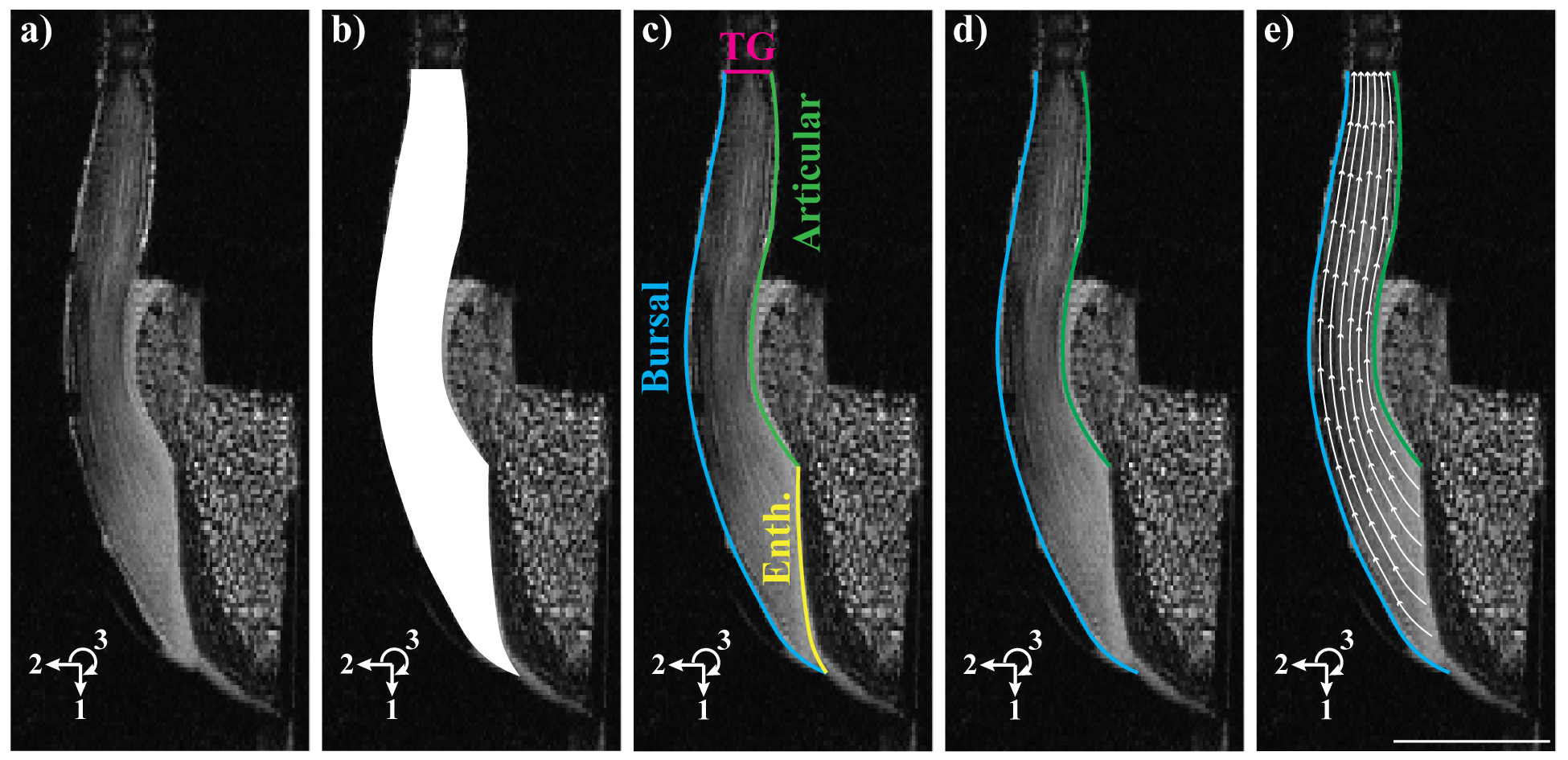}
    \caption{Procedure to find fiber directions in each 1-2 slice of high-resolution datasets. a) A representative 1-2 slice depicting the entire field of view captured by the MRI experiment. b) Binary mask isolates region of interest (tendon). c) Perimeter of mask shows bursal (cyan) and articular (green) surfaces, and tendon (magenta) and enthesis (yellow) boundaries. TG = tendon grip, Enth. = enthesis. d) Tendon and enthesis edges are removed. e) Fiber directions are found via interpolation from articular to bursal curves. Scale bar = 10 mm.}
    \label{fig:FiberDirections}
\end{figure}

A voxel-wise fiber material direction map, $\mathbf{a}_0 (\mathbf{X})$, was constructed based on an interpolation function between the curvatures of the articular and bursal edges of the tendon (Figure \ref{fig:FiberDirections}), using high-resolution anatomical slices in the 1-2 plane. To extract the curves that defined these edges in each slice, we first created a binary perimeter of the high-resolution masks with the MATLAB function \textit{bwmorph}. Then, several points were set to zero along the edges of the enthesis and tendon grip. These broken curves were eliminated using a threshold in the MATLAB function \textit{bwareaopen} so that only articular and bursal edges were kept. Edges of interest were extracted with the function \textit{bwboundaries}, and a fourth-order polynomial function was fit to each edge's cloud of points using the \textit{polyfit} function. A numerical derivative of the polynomial fit in each edge voxel was calculated with the function \textit{gradient}. To find the numerical derivatives of the internal regions, a linear interpolation was applied between the articular and bursal edges with the function \textit{inpaint\_nans}. The fiber direction components were derived from these numerical derivatives and normalized by their magnitude to define fiber direction unit vectors. It was assumed that there was no fiber misalignment in the 1-3 plane. A similar fiber direction generation procedure was used for the anterior cruciate ligament in the work by \citep{luetkemeyer2021constmod}.

\subsubsection{Mesh convergence}
\label{Sect:MethodsConvergence}

Five meshes were generated from each binary mask, with characteristic tetrahedral element sizes of 1 mm (mesh \#1, coarsest), 0.75 mm (mesh \#2), 0.5 mm (mesh \#3), 0.4 mm (mesh \#4), and 0.3 mm (mesh \#5, finest). Mesh quality, including aspect ratio and element shape, was assessed in Hypermesh. VSI inference using the modified HGO model was performed under four loading conditions for each mesh. To evaluate mesh convergence, we computed the displacement and load losses defined in Equation~\ref{Eqn:TotalLossVSI} using the parameters inferred for each mesh. These losses were plotted against the average number of elements (intact and torn combined) for each element size (Supplementary Figures \ref{fig:DisplacementLoss} and \ref{fig:LoadLoss}).

Based on this analysis, mesh \#3 (0.5 mm) provided a good balance between accuracy and computational efficiency and was selected for the remainder of the training and validation procedures. However, forward solutions using first-order tetrahedral elements, while yielding visually realistic displacements, produced strain fields with considerable noise. To address this, we adopted second-order tetrahedral elements in the coarsest mesh (mesh \#1), which better matched the resolution of the experimental displacement maps. This mesh was used for all subsequent inferences, as it produced comparable parameters to mesh \#3 while substantially reducing strain noise and computational cost (Supplementary Table \ref{tab:ComparisonCoarseFineMesh}).

\subsubsection{Dataset partitioning approach}
\label{Sect:MethodsPartition}
We divided each tendon's four datasets into training and validation sets, similar to an 80/20 partitioning scheme commonly used in machine learning. The intact condition with a 1 mm tensile displacement, applied longitudinally to specimens approximately 30 mm in length (see Figure \ref{fig:FiberDirections}-e for specimen scale), was chosen as the validation dataset. This dataset was selected because VSI benefits from experiments activating diverse deformation mechanisms, and the 1 mm intact condition, being at a low load and undamaged state, was considered the least information-rich. The remaining datasets - Intact-2 mm, Torn-1 mm, Torn-2 mm - were used for inference and training of tendon-specific models. The tensile mode of deformation was chosen to mimic the physiological loading of tendons at the neutral position, with elongations of 1 mm and 2 mm of elongation corresponding to submaximal and supramaximal shoulder force ranges, respectively, as stated in Section \ref{Sect:MethodsSamplePrep}.

\subsubsection{Voxel-wise error assessment}
\label{Sect:MethodsErrorMaps}
In total, 12 models were evaluated using full volume percentage error maps of displacements and invariants, in the training and validation datasets. The normal and shear strain components were excluded from the assessment, as the displacements were the input to our inference framework and already included in the evaluation, and the invariants constituted the basis of the strain energy density functions. The component-wise displacement error was calculated with the following equation,

\begin{equation}
    e_{u_j^i} = \frac{u_{j\text{FE}}^{i} - u_{\text{d}j}^\text{i}}{|\mathbf{u_{max}}|},
\end{equation}

where $u_j^i$ is the $i$-th component of the displacement in every $j$-th node of the mesh, the subscript FE indicates the forward model solution, the subscript d indicates the experimental data, and $\mathbf{u_{max}}$ is the maximum magnitude of the experimental displacement field. Given the possibility of finding close to zero nodal displacements, the maximum displacement magnitude was selected as the normalizing quantity.

To build the invariant error map, we used the following equation,

\begin{equation}
    e_{I_j^i} = \frac{I_{j\text{FE}}^i-I_{j\text{d}}^i}{I_{j\text{d}}^i},
\end{equation}

where $I_j^i$ is the $i$-th invariant of the right Cauchy-Green tensor in every $j$-th node of the mesh, and the subscripts FE and d have been described above.

The Torn-2 mm dataset was selected as the representative set to evaluate the calibration error, as we emphasized capturing the high shear feature in our models.

\subsubsection{Global error assessment}
\label{Sect:MethodsTotalError}

The squared volume-averaged $L^2_{2k}$-norm normalized with respect to the maximum displacement was calculated with the following equation,

\begin{equation}
    L_{2k}^2 = \frac{1}{V_k}\int_\Omega  \frac{\left|\left| \mathbf{u}_k^{FE} - \mathbf{u}_k \right|\right|^2}{||(\textbf{u}_{\text{max}})_k||^2} \text{d}v,
\end{equation}

where the subscript $k$ indicates that the error corresponds to individual training and validation sets.

\section{Results}
\label{Results}
\subsection{Parameter inference}
\label{Sect:ResultsParamInf}

Tables~\ref{tab:VSIvAdjoint_NH}, \ref{tab:VSIvAdjoint_HGO}, and \ref{tab:VSIvAdjoint_Polyn} present the inferred material parameters obtained from the VSI framework with parameter suppression, alongside those refined via PDE-constrained optimization, for the NH, m-HGO, and Pol models, respectively. Parameters suppressed by VSI have been omitted from the tables. In both the m-HGO and the Pol models, VSI suppressed the coefficients related to the higher-order invariants, indicating a reduced constitutive representation was sufficient to explain the data. Specifically, in the m-HGO model, the inclusion of $I_1$ within the exponential term led to negligible contribution from the higher-order terms to the stress response. In the Pol model, only three coefficients remained after parameter suppression, which indicated that $I_1$ dominated over $I_2$ and that a quadratic form of $I_4$ was sufficient to capture the fiber-related physics. Thus, under the present loading conditions, VSI was able to capture the dominant mechanics with a lower-dimensional parameter set than the full proposed models.

Beyond the suppressed parameters, additional trends were observed in the inferred bulk and shear moduli. In most tendons, VSI-based inference predicted a bulk modulus at the imposed lower boundary (0.2 MPa), indicating poor identifiability. In contrast, PDE-constrained refinement generally yielded bulk moduli in the near-incompressible regime. Across all models, the shear modulus generally decreased after refinement. In the m-HGO model, VSI often predicted perfectly aligned fibers ($\kappa \approx 0$), while PDE-constrained optimization introduced fiber dispersion in most tendons ($0<\kappa\leq 1/3$). The refinement also typically resulted in a reduction in the fiber stiffness parameter $k_1$ and the nonlinear fiber behavior parameter $k_2$. However, in certain tendons,  specifically tendons 1, 5, 6, 7, and 11, $k_1$ increased during refinement, indicating an enhanced fiber contribution. Additionally, tendons 5 and 6 experienced an increase in $k_2$, suggesting a more pronounced nonlinear fiber response.

For the polynomial model, the fiber-related coefficient - which would be analogous to $k_1$ - generally decreased following refinement, consistent with the trends observed in the m-HGO formulation.

\begin{table}[h!]
    \centering
    {\small
    \renewcommand{\arraystretch}{1.2} 
    \setlength{\tabcolsep}{5pt} 
    \begin{tabular}{c c c c c}
    \hline
        \multicolumn{5}{c}{Candidate model, $W_{\text{NH}}$} \\
        \hline
        \multirow{3}{*}{Tendon} & \multicolumn{2}{c}{VSI} & \multicolumn{2}{c}{Adjoint} \\
        \cline{2-5}
        & $K$ & $\mu$ & $K$ & $\mu$\\
        & (MPa) & (MPa) & (MPa) & (x$10^{-2}$ MPa) \\
        \hline
        1 & 0.20 & 1.48 & 767.18 & 7.44 \\
        2 & 0.20 & 1.14 & 677.46 & 5.72 \\
        3 & 5.30 & 3.28 & 319.14 & 1.34 \\
        4 & 0.20 & 0.26 & 896.94 & 6.54 \\
        5 & 0.24 & 1.06 & 98.02 & 1.06 \\
        6 & 0.26 & 2.34 & 394.4 & 11.22 \\
        7 & 4.52 & 0.80 & 231.78 & 1.94 \\
        8 & 0.20 & 1.06 & 268.22 & 1.06 \\
        9 & 0.24 & 2.88 & 368.44 & 4.50 \\
        10 & 0.20 & 0.08 & 42.78 & 0.4 \\
        11 & 2.14 & 0.78 & 68.5 & 0.78 \\
        12 & 0.20 & 0.48 & 266.56 & 1.86 \\
    \hline
    \end{tabular}}
    \caption{Parameters inferred for the NH model, using the VSI framework and PDE-constrained refinement.}
    \label{tab:VSIvAdjoint_NH}
\end{table}

\begin{table}[h!]
    \centering
    {\footnotesize
    \renewcommand{\arraystretch}{1.2} 
    \setlength{\tabcolsep}{2pt} 
    \begin{tabular}{c c c c c c c c c c c}
        \hline
        \multicolumn{11}{c}{Candidate model, $W_{\text{m-HGO}}$} \\
        \hline
        \multirow{3}{*}{Tendon} & \multicolumn{5}{c}{VSI} & \multicolumn{5}{c}{Adjoint} \\
        \cline{2-11}
        & $K$ & $\mu$ & $k_1$ & $k_2$ & \multirow{2}{1em}{$\kappa$} & $K$ & $\mu$ & $k_1$ & $k_2$ & \multirow{2}{1em}{$\kappa$} \\
        & (MPa) & (MPa) & (MPa) & (x$10^{-1}$) & & (MPa) & (x$10^{-2}$ MPa) & (MPa) & (x$10^{-1}$) & \\
        \hline
        1 & 0.20 & 1.08 & 1.40 & 6.84 & 0 & 238.84 & 9.70 & 19.70 & 0.23 & 0.22 \\
        2 & 0.20 & 0.38 & 1.42 & 2.65 & 0 & 597.62 & 5.74 & 0.01 & 0.16 & 0.33 \\
        3 & 1.70 & 1.36 & 3.62 & 1.23 & 0 & 249.6 & 1.36 & 0.29 & 0.99 & 0.18 \\
        4 & 0.20 & 0.06 & 0.36 & 5.96 & 0 & 251.14 & 9.60 & 0.90 & 0.59 & 0 \\
        5 & 0.20 & 0.20 & 3.81 & 0.14 & 0 & 352.30 & 2.32 & 16.04 & 66.49 & 0.32 \\
        6 & 0.20 & 0.76 & 5.22 & 11.13 & 0.08 & 313.9 & 16.54 & 63.60 & 11.32 & 0.28 \\
        7 & 1.38 & 0.30 & 3.54 & 0.07 & 0.10 & 4.64 & 10.3 & 3.39 & 0.84 & 0.17 \\
        8 & 0.20 & 0.40 & 3.30 & 5.24 & 0 & 705.2 & 1.74 & 0.19 & 4.05 & 0.03 \\
        9 & 0.20 & 1.90 & 2.50 & 0.13 & 0 & 368.9 & 1.90 & 0.57 & 0.13 & 0.24 \\
        10 & 0.20 & 0.08 & 0.12 & 3.04 & 0 & 100.02 & 5.68 & 0.10 & 3 & 0.02 \\
        11 & 0.70 & 0.24 & 1.25 & 0.16 & 0.002 & 14 & 49.22 & 2.23 & 0.15 & 0 \\
        12 & 0.20 & 0.44 & 0.70 & 8.34 & 0.27 & 801.8 & 6.60 & 0.35 & 4.17 & 0.14 \\
        \hline
    \end{tabular}}
    \caption{Parameters inferred for the m-HGO model, using the VSI framework and PDE-constrained refinement.}
    \label{tab:VSIvAdjoint_HGO}
\end{table}

\begin{table}[h!]
    \centering
    {\small
    \renewcommand{\arraystretch}{1.2} 
    \setlength{\tabcolsep}{5pt} 
    \begin{tabular}{c c c c c c c}
        \hline
        \multicolumn{7}{c}{Candidate model, $W_{\text{m-Pol}}$} \\
        \hline
        \multirow{3}{*}{Tendon} & \multicolumn{3}{c}{VSI} & \multicolumn{3}{c}{Adjoint} \\
        \cline{2-7}
        & $K$ & $\mu$ & $\theta_7$ & $K$ & $\mu$ & $\theta_7$ \\
        & (MPa) & (MPa) & (MPa) & (MPa) & (x$10^{-2}$ MPa) & (MPa) \\
        \hline
        1 & 0.20 & 0.96 & 0.97 & 291.12 & 3.32 & 0.24 \\
        2 & 0.20 & 0.30 & 0.79 & 506.62 & 30.84 & 0.39 \\
        3 & 1.78 & 1.46 & 2.07 & 298.68 & 1.46 & 0.04 \\
        4 & 0.20 & 0.08 & 0.22 & 277.96 & 4.64 & 0.24 \\
        5 & 0.20 & 0.12 & 1.91 & 290.12 & 11.34 & 1.53 \\
        6 & 0.20 & 0.50 & 2.28 & 573.26 & 19.58 & 0.34 \\
        7 & 1.90 & 0.34 & 1.12 & 28.44 & 2.82 & 0.09 \\
        8 & 0.20 & 0.32 & 1.86 & 641.74 & 6.72 & 0.33 \\
        9 & 0.20 & 1.82 & 1.27 & 8.34 & 107.32 & 1.18 \\
        10 & 0.20 & 0.08 & 0.15 & 40 & 16 & 0.12 \\
        11 & 0.70 & 0.24 & 0.62 & 136.48 & 24.10 & 0.50 \\
        12 & 0.20 & 0.44 & 0.10 & 141.46 & 8.68 & 0.08\\
        \hline
    \end{tabular}}
    \caption{Parameters inferred for the Pol model, using the VSI framework and PDE-constrained refinement.}
    \label{tab:VSIvAdjoint_Polyn}
\end{table}

\subsection{Full volume maps from forward predictions}
\label{Sect:ResultsErrorMaps}

Forward predictions were obtained using the parameters from VSI and PDE-constrained optimizations, respectively, for all tendons. Although a global $L_2$-norm was used for parameter optimization, full volume displacement and invariant errors were calculated to observe regions of improvement. Our full volume results are explained for a representative tendon (tendon 1) in the following sections.

\subsubsection{Displacement and displacement error maps}
\label{Sect:DispMaps}

\begin{figure}[h!]
    \centering
    \includegraphics[width=\linewidth]{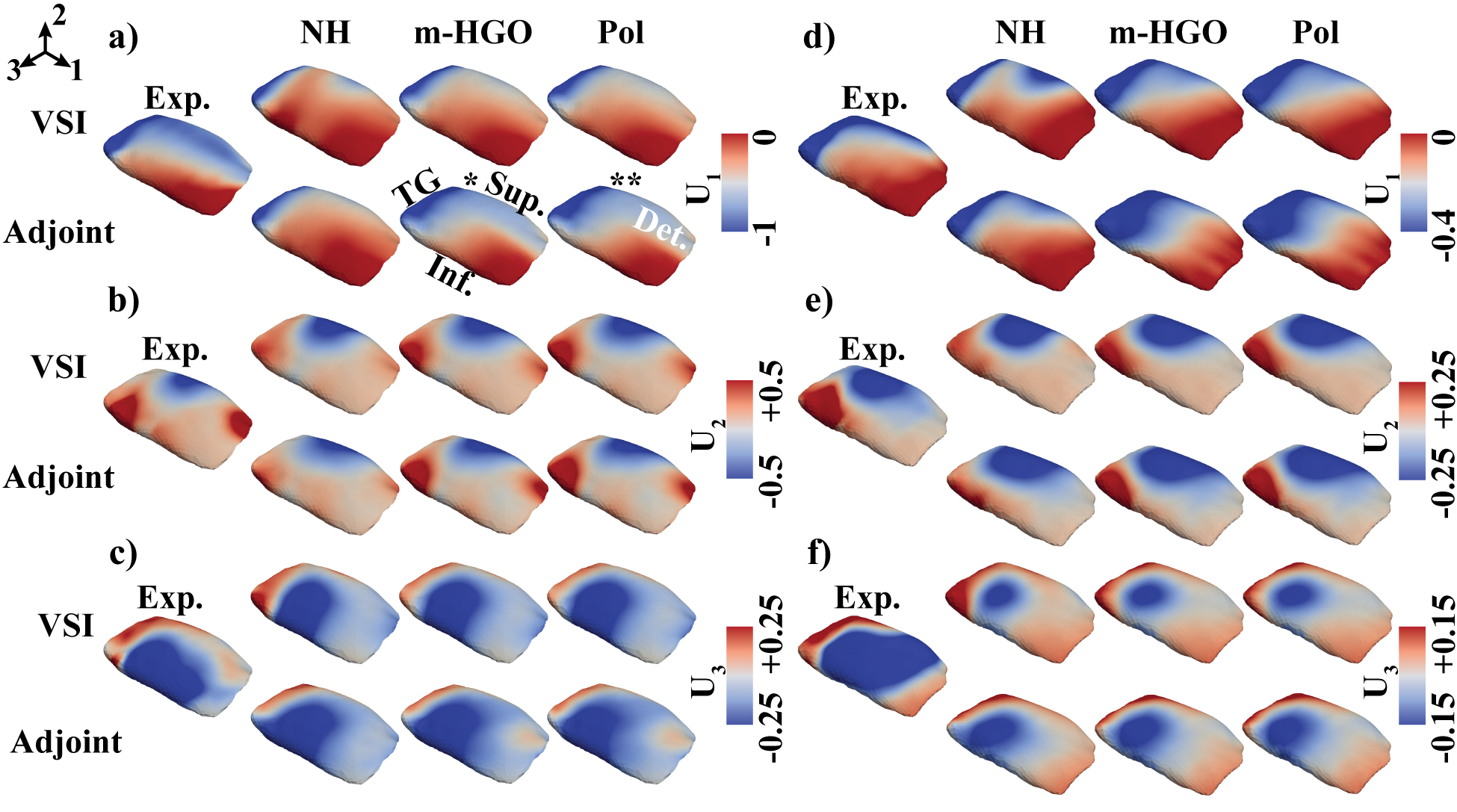}
    \caption{Full volume displacement maps of a representative tendon (right shoulder). Each frame depicts the experimental response on the left, accompanied by forward predictions from VSI inference (first row) and adjoint refinement (second row). Displacements in the longitudinal (a, d), thickness (b, e) and transverse (c, f) direction correspond to the torn condition at the 2 mm elongation (a, b, c) and the intact condition at the 1 mm elongation (validation dataset, frames d, e, f), respectively. *TG = tendon gripper, Sup. = superior side, Inf. = inferior side. **Det. = detached tissue band.}
    \label{fig:DisplacementComparison}
\end{figure}

\begin{figure}[h!]
    \centering
    \includegraphics[width=\linewidth]{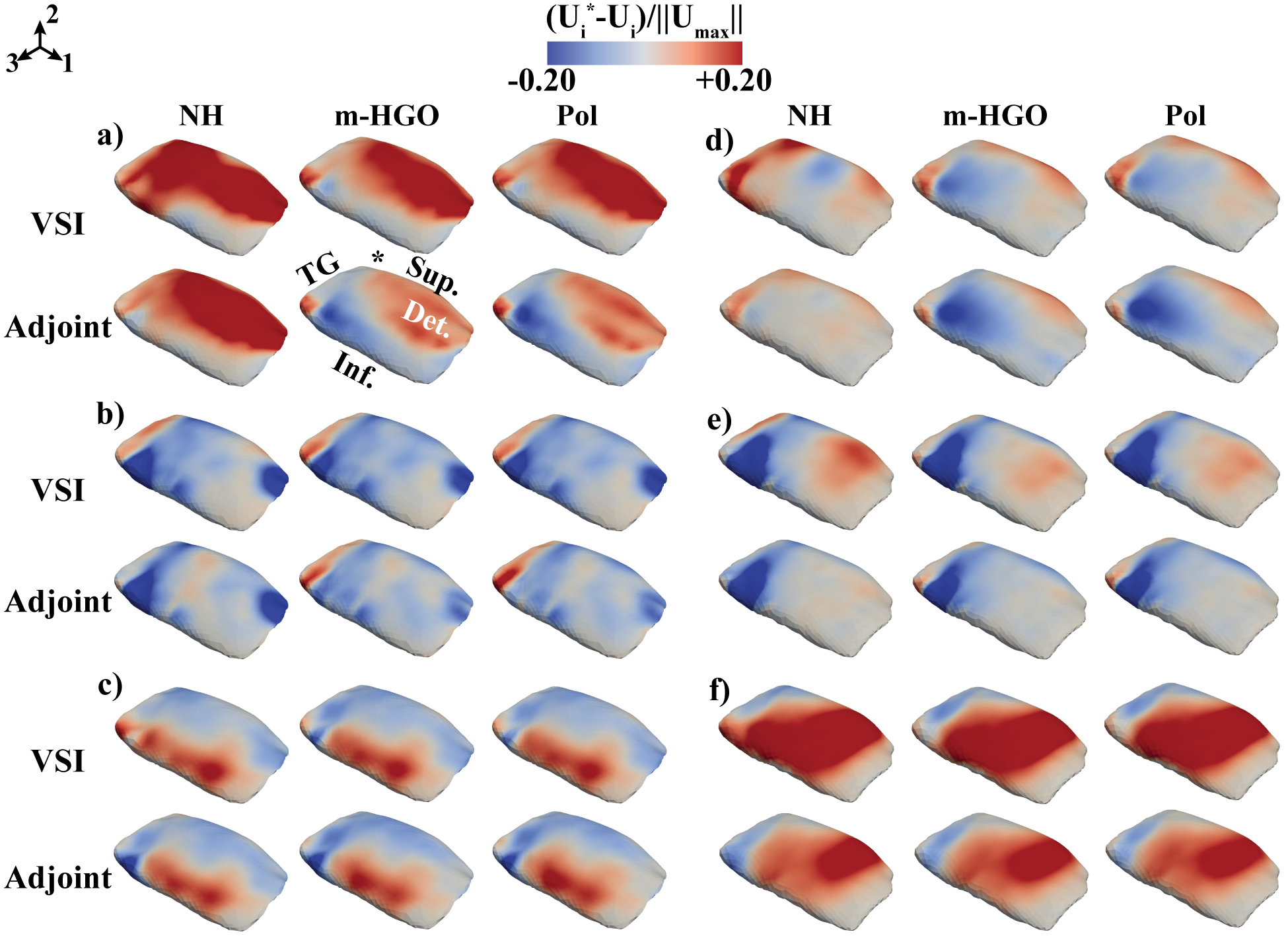}
    \caption{Full volume displacement error maps of a representative tendon (right shoulder). Each frame depicts the forward prediction errors from VSI inference (first row) and adjoint refinement (second row). Error displacements in the longitudinal (a, d), thickness (b, e) and transverse (c, f) direction correspond to the torn condition at the 2 mm elongation (a, b, c) and the intact condition at the 1 mm elongation (validation dataset, frames d, e, f), respectively. *TG = tendon gripper, Sup. = superior side, Inf. = inferior side, Det. = detached tissue band.}
    \label{fig:DisplacementErrorComparison}
\end{figure}

Using parameters estimated via VSI, the NH model had the poorest performance of the three predictions, when considering the torn configuration at the supramaximal load (Figure \ref{fig:DisplacementComparison}-a to c, first row). While the characteristic sliding behavior in the longitudinal direction observed in the torn condition was partially captured by all models (Figure \ref{fig:DisplacementComparison}-a), the NH model displayed larger errors in the detached band of tissue (Figure \ref{fig:DisplacementErrorComparison}-a) and close to the tendon gripper (Figure \ref{fig:DisplacementErrorComparison}-a, b). All models exhibited localized regions where displacement errors exceeded 20\% in all tendons, particularly in the detached band of tissue for the longitudinal displacement (Figure \ref{fig:DisplacementErrorComparison}-a), and the inferior side of the tendon for the transverse displacement (Figure \ref{fig:DisplacementErrorComparison}-c). The Pol and m-HGO models produced similar displacement and error maps (Figures \ref{fig:DisplacementComparison} and \ref{fig:DisplacementErrorComparison}), which could be explained due to the similar magnitude of the fiber stiffness parameter, $k_1$ in the m-HGO model and $\theta_7$ in the Pol model (Tables \ref{tab:VSIvAdjoint_HGO} and \ref{tab:VSIvAdjoint_Polyn}). In general, errors in the transverse and thickness directions were smaller in magnitude as compared to the longitudinal direction (Figure \ref{fig:DisplacementErrorComparison}-b and c), emphasizing the importance of improving longitudinal displacement predictions. The PDE-constrained refined parameters yielded more accurate displacement maps in all directions (Figure \ref{fig:DisplacementComparison}-a to c, second row), with the m-HGO and Pol models producing an increased sliding behavior. While the adjoint-refined NH model presented reduced error near the tendon gripper (Figure \ref{fig:DisplacementErrorComparison}-a) as compared to the VSI prediction, improvements to capture larger tissue sliding were not achieved. The refined m-HGO and Pol models significantly reduced the error in the detached tissue band, in the longitudinal direction (Figure \ref{fig:DisplacementErrorComparison}-a), and were able to reproduce the positive displacement in the thickness direction at the detached footprint (Figure \ref{fig:DisplacementErrorComparison}-b). A small reduction of transverse error was observed in the inferior side of the tendon (Figure \ref{fig:DisplacementErrorComparison}-c), particularly for the m-HGO model.

At the submaximal load in the intact configuration (our validation set), the Pol and m-HGO models more accurately captured the longitudinal displacement gradient between the superior and inferior regions of the tendon (Figure \ref{fig:DisplacementComparison}-d) using VSI-inferred parameters. In contrast, the NH model exhibited a more uniform distribution of longitudinal displacements, failing to reproduce the observed spatial gradient (Figure \ref{fig:DisplacementComparison}-d). This led to generally larger longitudinal error magnitudes in the NH model, particularly on the superior side of the tendon. We observed increased error values in the transverse and thickness directions as compared to the longitudinal direction, with transverse displacement errors larger than 20\% in the midsubstance, covering a significant portion of the tendon length (Figure \ref{fig:DisplacementErrorComparison}-f). In the thickness direction, the m-HGO and Pol predictions generally achieved smaller errors in the area close to the footprint when measured against the NH prediction (Figure \ref{fig:DisplacementErrorComparison}-e). With the PDE-constrained refinement, longitudinal displacement errors decreased for the NH model (Figure \ref{fig:DisplacementErrorComparison}-a), indicating large dependence on incompressibility to accurately capture realistic behavior. In the thickness direction, all models closely followed the experimental maps in the area close to the footprint and the superior side of the tendon, with the m-HGO and Pol models capturing the displacement gradient near the tendon gripper (Figure \ref{fig:DisplacementComparison}-e). The largest disagreement of all models was in the transverse direction, but reduced errors were found with the refined parameters (Figure \ref{fig:DisplacementErrorComparison}-f), which localized error magnitudes near the curved portion of the tendon at the bursal surface.

Importantly, when considering all tendons, tendon 12 had the highest fiber dispersion value (Table \ref{tab:VSIvAdjoint_HGO}), which led to decreased transverse errors in the HGO prediction relative to the Pol prediction (not shown) for VSI-inferred parameters. Overall, after the application of parameter refinement, voxel-wise errors consistently decreased in the longitudinal direction, but were localized in the detached tissue portion in tendons 2, 6, and 7. In the thickness direction, errors generally became more localized after PDE-constrained refinement. Errors in the transverse direction were the largest for the validation dataset, with larger error concentration zones than the selected inference set, similar to tendon 1 (the representative tendon).

\subsubsection{Strain maps}
\label{Sect:DispErrorMaps}

\begin{figure}[h!]
    \centering
    \includegraphics[width=\linewidth]{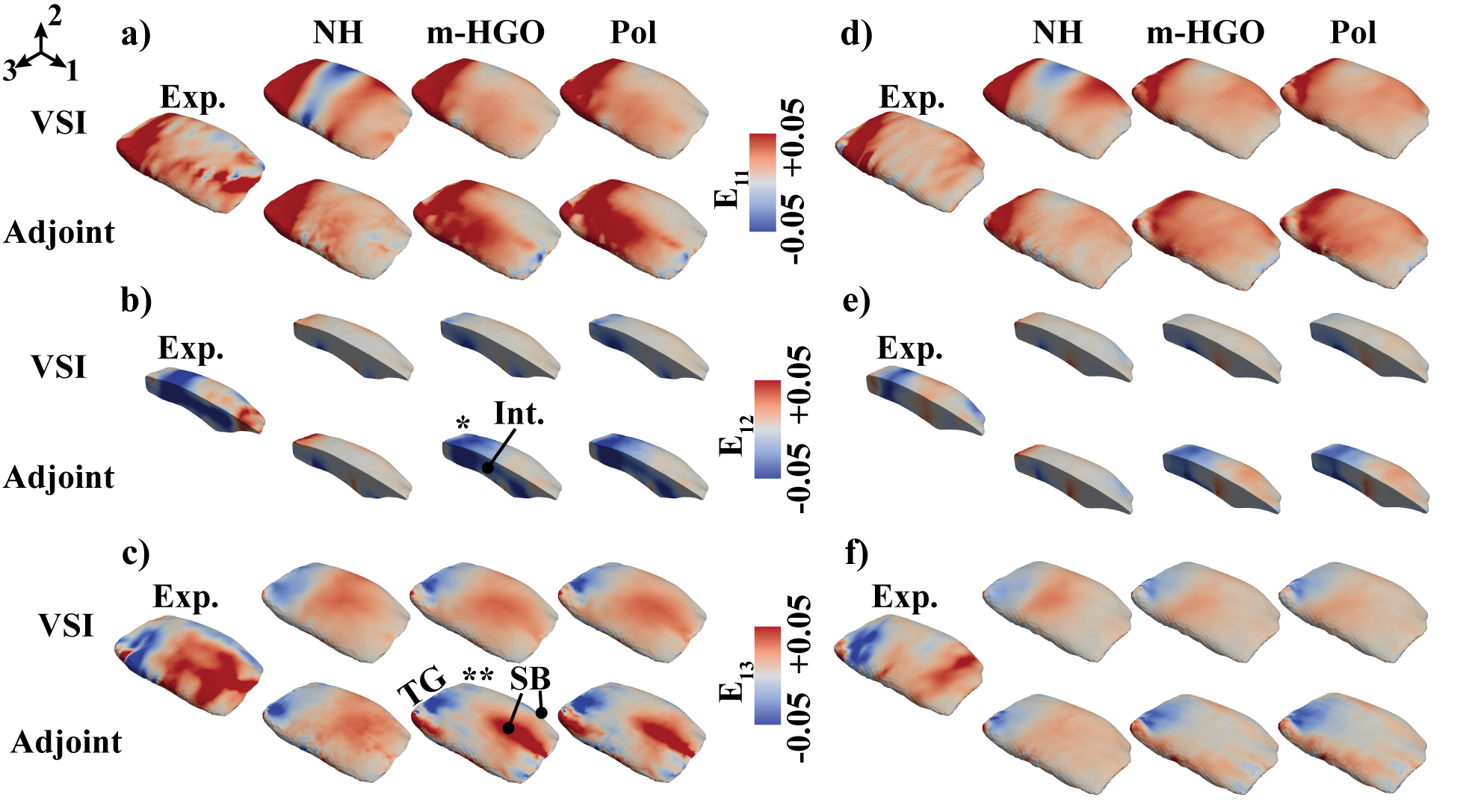}
    \caption{Full volume strain maps of a representative tendon. Each frame depicts the experimental response on the left, accompanied by forward predictions from VSI inference (first row) and adjoint refinement (second row). Longitudinal (a, d), 1-2 shear (b, e) and 1-3 shear (c, f) strain correspond to the torn condition at the 2 mm elongation (a, b, c) and the intact condition at the 1 mm elongation (validation dataset, frames d, e, f), respectively. Note how the m-HGO and Pol models capture internal strain variation after PDE-constrained optimization (b, e). *Int. = internal region of tendon. **TG = tendon gripper, SB = shear bands at the cut-uncut tissue interface.}
    \label{fig:StrainComparison}
\end{figure}

When considering the strain maps of the torn condition at the 2 mm elongation from the VSI-based inference, all models were able to reproduce the high longitudinal strain near the tendon gripper (Figure \ref{fig:StrainComparison}-a). However, the NH model predicted an unrealistic strain distribution, characterized by a sequence of tensile, compressive, and tensile strain regions along the tendon length. This non-physiological compressive strain region, not observed in experimental datasets, could have arisen due to the lack of model anisotropy. The m-HGO and Pol models slightly captured the internal 1-2 shear strain band (Figure \ref{fig:StrainComparison}-b). The 1-3 shear strain was similar for all models, without displaying shear strain concentrations like the experimental response (Figure \ref{fig:StrainComparison}-c). After the PDE-constrained optimization, the NH model no longer displayed the tension-compression-tension gradient in the longitudinal strain, and m-HGO and Pol models localized longitudinal strains near the inferior side of the tendon (Figure \ref{fig:StrainComparison}-a). While m-HGO and Pol models clearly captured the internal 1-2 shear strain localization, the NH model was not able to reproduce it after optimization (Figure \ref{fig:StrainComparison}-b). Similarly, the two characteristic 1-3 shear strain concentration bands at the inferior and superior regions were shown in the m-HGO and Pol models, but not the NH model (Figure \ref{fig:StrainComparison}-c).

The VSI-based forward strain predictions of the validation set displayed similar behavior as compared to the selected inference set (Figure \ref{fig:StrainComparison}-d to f). After the optimization, improvements were observed in the 1-2 shear strain component for the m-HGO and Pol models (Figure \ref{fig:StrainComparison}-e), specifically, these models reproduced the shear strain gradient between distal and proximal portions of the tendon. In the 1-3 shear strain component, refined m-HGO and Pol models slightly captured strain concentration near the tendon gripper (Figure \ref{fig:StrainComparison}-f) and the inferior side of the tendon.

When considering all tendons, the optimized NH models consistently failed to accurately reproduce internal 1-2 shear strains and 1-3 shear strain bands which were key for the identification of delamination in experimental torn tendons. 

\subsubsection{Invariant error maps}
\label{Sect:InvariantMaps}

\begin{figure}[h!]
    \centering
    \includegraphics[width=\linewidth]{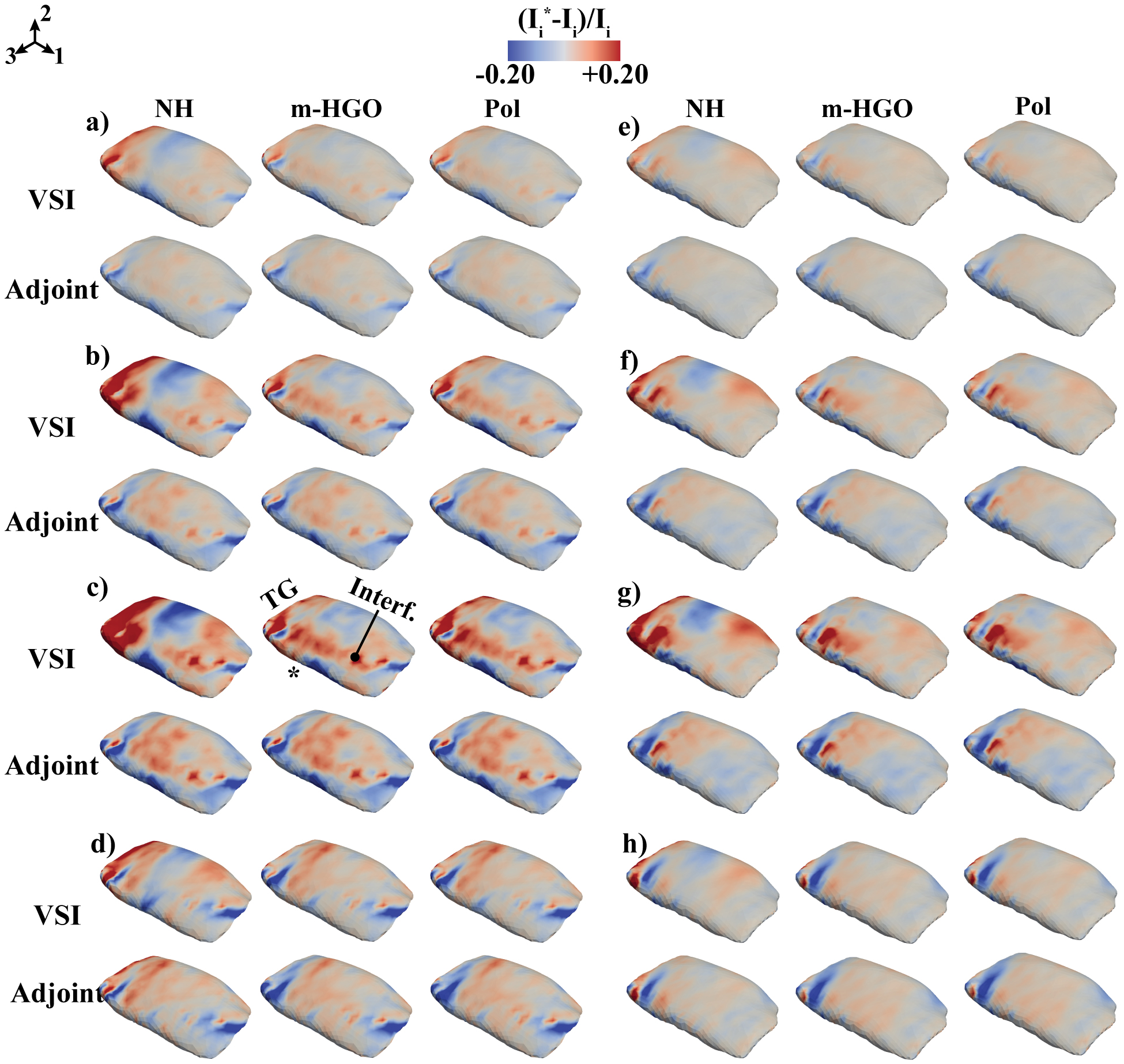}
    \caption{Full volume invariant error maps of a representative tendon. Each frame depicts the forward prediction errors from VSI inference (first row) and adjoint refinement (second row). Invariant errors for the first, $I_1$, (a, e), second, $I_2$, (b, f), third, $I_3$, (c, g) and fourth, $I_4$, (d, h) invariants correspond to the torn condition at the 2 mm elongation (frames a to d) and the intact condition at the 1 mm elongation (validation dataset, frames e to h), respectively. *TG = tendon gripper, Interf. = interface between detached and attached tissue portions.}
    \label{fig:InvariantErrorComparison}
\end{figure}

In the torn/supramaximal combination of VSI-inferred predictions, all $I_2$ and $I_3$ error maps displayed the largest concentrations (Figures \ref{fig:InvariantErrorComparison}-b and c), with values above 50\%. The area near the tendon gripper presented larger errors for all the invariants in the NH model as compared to the m-HGO and Pol models (Figure \ref{fig:InvariantErrorComparison}). $I_2$ and $I_3$ error concentration areas were found at the cut/uncut tissue interface and tendon gripper (Figure \ref{fig:InvariantErrorComparison}-c). $I_4$ errors were localized near the footprint, at the detached area (Figure \ref{fig:InvariantErrorComparison}-d). With refinement, all errors considerably decreased, particularly for $I_2$ and $I_3$, in all models (Figure \ref{fig:InvariantErrorComparison}-a to d, second row). For $I_4$, errors persisted near the attachment, which could indicate that the representation of the fiber direction field could be redefined for an improved prediction.

Invariant error maps $I_1$, $I_2$, and $I_4$ from VSI-inferred predictions showed values generally smaller than 10\% at the submaximal load in the intact configuration (Figure \ref{fig:InvariantErrorComparison}-e, f, and h). The largest source of error was found in $I_3$, close to the tendon gripper, particularly for the NH model (Figure \ref{fig:InvariantErrorComparison}-g). After refinement, all maps except for $I_4$ showed decreased error localization. In particular, the error decrease in $I_3$, which is directly related to the volumetric deformation, highlights how incompressibility improved the predictions. As with the torn condition at the supramaximal load, the fiber direction field could be improved to decrease $I_4$ errors near the tendon gripper (Figure \ref{fig:InvariantErrorComparison}-h). 

Analyzing all tendons, noisy $I_4$ regions were observed at the inferior edges of tendons 2 and 11 (not shown). Some tendons displayed a $I_4$ error gradient between the area close to the enthesis and the area close to the gripper, again highlighting the need for an improved fiber direction field. $I_3$ error was concentrated at the cut/uncut tissue boundaries, near the tendon gripper, and the enthesis, as with the representative tendon. The NH prediction developed larger errors as compared to the Pol and m-HGO predictions, in all cases. 

\subsection{Global error comparison between VSI and adjoint inference}
\label{Sect:GlobalErrorComparison}

Tables \ref{tab:GlobalErrorVSIAdjointNH}, \ref{tab:GlobalErrorVSIAdjointHGO}, and \ref{tab:GlobalErrorVSIAdjointPol} report the squared $L_2$ norms for the NH, m-HGO, and Pol models, respectively, comparing VSI- and adjoint-derived predictions. For all models, the majority of tendons showed a reduction in $L_2$-error norm of the solution obtained from the VSI-inferred model after parameter refinement on the selected inference dataset. An exception was tendon 10, which had an increased squared $L_2$ norm for both the NH and m-HGO models, suggesting that other inference datasets were more influential during optimization. For the validation dataset, tendons 3 and 11 showed an increase in the squared $L_2$ norm for the optimized NH model. Similarly, tendons 3, 7, and 11 had increased validation error norms with the refined m-HGO model. The optimized Pol model showed higher validation error norms in tendons 3, 10, and 11. Importantly, an increased $L_2$ error may be influenced by localized areas with high contribution to the global error.

Comparing the resulting squared $L_2$ norms of all refined models in the selected training dataset, the m-HGO and Pol models reported the smallest values for tendons 1, 6, 8, 11 and 12, and 3, 4, 5, 7 and 9, respectively (text in bold in Tables \ref{tab:GlobalErrorVSIAdjointHGO} and \ref{tab:GlobalErrorVSIAdjointPol}). Only two tendons (2, 10) showed improved fit with the refined NH model (Table \ref{tab:GlobalErrorVSIAdjointNH}). This highlights the importance of including an anisotropic term in the strain energy density function to accurately capture all deformation mechanisms. For the validation datasets, the smallest errors were found primarily in the m-HGO model (6 tendons), followed by the Pol model (4 tendons), and the NH model (2 tendons).

\begin{table}[h!]
    \centering
    {\small
    \renewcommand{\arraystretch}{1.2} 
    \setlength{\tabcolsep}{5pt} 
    \begin{tabular}{c c c c c c c}
        \hline
        \multicolumn{7}{c}{Candidate model, $W_{\text{m-NH}}$} \\
        \hline
        \multirow{4}{*}{Tendon} & \multicolumn{3}{c}{Selected training dataset} & \multicolumn{3}{c}{Validation dataset}\\
        & \multicolumn{3}{c}{Torn, 2 mm} & \multicolumn{3}{c}{Intact, 1 mm}\\
        \cline{2-7}
        & VSI $L_2^2$ & Adjoint $L_2^2$ & \multirow{2}{*}{\% change} & VSI $L_2^2$ & Adjoint $L_2^2$ & \multirow{2}{*}{\% change} \\
        & $\left(\text{x}10^{-2}\right)$ & $\left(\text{x}10^{-2}\right)$ &  & $\left(\text{x}10^{-2}\right)$ & $\left(\text{x}10^{-2}\right)$ & \\
        \hline
        1 & 3.11 & 2.53 & (18.63) & 2.58 & 2.17 & (15.87) \\
        2 & 5.10 & \textbf{4.26} & (16.48) & 10.04 & \textbf{9.44} & (5.98) \\
        3 & 4.28 & 4.11 & (3.92) & 3.49 & 3.96 & 13.44 \\
        4 & 2.18 & 1.91 & (12.98) & 1.68 & 1.52 & (9.28) \\
        5 & 3.23 & 3.11 & (3.56) & 3.44 & 3.27 & (5.14) \\
        6 & 4.53 & 3.00 & (33.83) & 2.93 & 2.08 & (28.94) \\
        7 & 6.65 & 6.42 & (3.45) & 2.73 & \textbf{2.68} & (1.90) \\
        8 & 5.86 & 3.73 & (36.34) & 3.34 & 1.95 & (41.61) \\
        9 & 5.72 & 2.42 & (5.68) & 7.75 & 4.64 & (40.05) \\
        10 & 1.01 & \textbf{1.16} & 14.43 & 4.75 & 4.43 & (6.68) \\
        11 & 1.50 & 1.46 & (3.00) & 4.01 & 4.02 & 6.77 \\
        12 & 1.65 & 1.26 & (23.79) & 4.12 & 3.03 & (26.43) \\
        \hline
    \end{tabular}}
    \caption{Comparison between squared $L_2$ norm obtained with parameters inferred from VSI and PDE-constrained refinement, for the NH model. Values in bold indicate the smallest result among all models. Values in parenthesis represent a percentage decrease.}
    \label{tab:GlobalErrorVSIAdjointNH}
\end{table}

\begin{table}[h!]
    \centering
    {\small
    \renewcommand{\arraystretch}{1.2} 
    \setlength{\tabcolsep}{5pt} 
    \begin{tabular}{c c c c c c c}
        \hline
        \multicolumn{7}{c}{Candidate model, $W_{\text{m-HGO}}$} \\
        \hline
        \multirow{4}{*}{Tendon} & \multicolumn{3}{c}{Selected training dataset} & \multicolumn{3}{c}{Validation dataset}\\
        & \multicolumn{3}{c}{Torn, 2 mm} & \multicolumn{3}{c}{Intact, 1 mm}\\
        \cline{2-7}
        & VSI $L_2^2$ & Adjoint $L_2^2$ & \multirow{2}{*}{\% change} & VSI $L_2^2$ & Adjoint $L_2^2$ & \multirow{2}{*}{\% change} \\
        & $\left(\text{x}10^{-2}\right)$ & $\left(\text{x}10^{-2}\right)$ &  & $\left(\text{x}10^{-2}\right)$ & $\left(\text{x}10^{-2}\right)$ & \\
        \hline
        1 & 2.02 & \textbf{1.03} & (48.97) & 2.15 & 1.77 & (17.43) \\
        2 & 13.53 & 4.28 & (68.40) & 18.61 & 9.47 & (49.14) \\
        3 & 4.01 & 3.80 & (5.33) & 2.39 & 2.64 & 10.53 \\
        4 & 1.06 & 0.94 & (11.80) & 0.98 & 0.93 & (5.39) \\
        5 & 6.24 & 2.91 & (53.35) & 7.59 & 3.57 & (53.01) \\
        6 & 2.49 & \textbf{1.85} & (25.81) & 2.15 & \textbf{1.66} & (22.86) \\
        7 & 3.94 & 3.85 & (2.24) & 3.01 & 3.03 & 0.70 \\
        8 & 1.86 & \textbf{1.15} & (37.94) & 2.11 & \textbf{1.43} & (32.28) \\
        9 & 2.59 & 1.92 & (26.03) & 5.13 & \textbf{3.91} & (23.70) \\
        10 & 1.06 & 1.17 & 10.39 & 4.10 & \textbf{4.05} & (1.20) \\
        11 & 0.86 & \textbf{0.84} & (2.04) & 3.50 & \textbf{3.61} & 3.03 \\
        12 & 1.16 & \textbf{0.90} & (22.38) & 3.40 & \textbf{2.63} & (22.52) \\
        \hline
    \end{tabular}}
    \caption{Comparison between squared $L_2$ norm obtained with parameters inferred from VSI and PDE-constrained refinement, for the m-HGO model. Values in bold indicate the smallest result among all models. Values in parenthesis represent a percentage decrease.}
    \label{tab:GlobalErrorVSIAdjointHGO}
\end{table}

\begin{table}[h!]
    \centering
    {\small
    \renewcommand{\arraystretch}{1.2} 
    \setlength{\tabcolsep}{5pt} 
    \begin{tabular}{c c c c c c c}
        \hline
        \multicolumn{7}{c}{Candidate model, $W_{\text{Pol}}$} \\
        \hline
        \multirow{4}{*}{Tendon} & \multicolumn{3}{c}{Selected training dataset} & \multicolumn{3}{c}{Validation dataset}\\
        & \multicolumn{3}{c}{Torn, 2 mm} & \multicolumn{3}{c}{Intact, 1 mm}\\
        \cline{2-7}
        & VSI $L_2^2$ & Adjoint $L_2^2$ & \multirow{2}{*}{\% change} & VSI $L_2^2$ & Adjoint $L_2^2$ & \multirow{2}{*}{\% change} \\
        & $\left(\text{x}10^{-2}\right)$ & $\left(\text{x}10^{-2}\right)$ &  & $\left(\text{x}10^{-2}\right)$ & $\left(\text{x}10^{-2}\right)$ & \\
        \hline
        1 & 1.80 & 1.13 & (37.13) & 2.06 & \textbf{1.72} & (16.12) \\
        2 & 16.00 & 8.88 & (44.49) & 21.93 & 13.97 & (36.31) \\
        3 & 4.01 & \textbf{3.74} & (6.73) & 2.36 & \textbf{2.52} & 7.04 \\
        4 & 1.08 & \textbf{0.92} & (14.52) & 1.00 & \textbf{0.91} & (8.58) \\
        5 & 7.37 & \textbf{2.93} & (60.25) & 10.54 & \textbf{2.66} & (74.76) \\
        6 & 2.92 & 2.22 & (23.80) & 2.19 & 1.79 & (18.34) \\
        7 & 3.77 & \textbf{3.77} & (0.03) & 3.06 & 3.00 & (2.05) \\
        8 & 1.95 & 1.23 & (36.83) & 2.07 & 1.55 & (15.24) \\
        9 & 2.54 & \textbf{1.91} & (24.66) & 5.04 & 4.21 & (16.35) \\
        10 & 1.77 & 1.17 & (33.87) & 3.94 & 4.14 & 4.92 \\
        11 & 0.86 & 8.49 & (1.29) & 3.49 & 3.62 & 3.62 \\
        12 & 1.11 & 0.92 & (17.04) & 3.63 & 2.78 & (23.48) \\
        \hline
    \end{tabular}}
    \caption{Comparison between squared $L_2$ norm obtained with parameters inferred from VSI and PDE-constrained refinement, for the Pol model. Values in bold indicate the smallest result among all models. Values in parenthesis represent a percentage decrease.}
    \label{tab:GlobalErrorVSIAdjointPol}
\end{table}

\section{Discussion and Conclusion}
\label{Sect:DiscConc}

The objective of this study was to develop and assess tendon-specific constitutive models for ovine rotator cuff tendons using full-volume MRI datasets, which capture material heterogeneity, complex tendon geometry, and curved boundary conditions. This was achieved through a VSI framework coupled with a PDE-constrained optimization approach. A full-volume fiber direction field, aligned with the tendon natural curvature as it wrapped around the humeral head, was incorporated into the formulation. Three constitutive models were calibrated, and their corresponding training and validation errors were quantified. The following paragraphs highlight the key findings and implications of these model inferences.

\subsection{Full volume MRI enabled inverse modeling of tendon heterogeneity via VSI}
\label{Sect:DiscFullVolumeMRI}

The ability of our full volume MRI-based data to capture the activation of multiple deformation mechanisms in both intact and torn tendons provides a unique opportunity to leverage inverse characterization methods such as VSI and its adjoint refiner. With a single quasi-uniaxial loading test, we observed a combination of mechanical responses, including shear, localized tension, and compression along different directions. 

Importantly, our parameter inference approach based on full volume datasets accounts for the global tendon behavior and incorporates the influence of native geometry and internal structure on three-dimensional deformation. This framework provides a clear advantage over calibration methods relying on surface strain, which overlook through-thickness information that we previously demonstrated to be mechanically meaningful \citep{luetkemeyer2021constmod, villacis2025mri}. Furthermore, our method avoids the need to physically excise the tendon into multiple regions to calibrate distinct regional properties. This is particularly valuable, as it suggests that tendons can be modeled as continuous structures, with internal mechanical heterogeneity and strain trends effectively being captured by traditional constitutive models. Our findings appear to support a more unified modeling approach for the rotator cuff, contrary to the emerging trend of regional subdivision \citep{garcia2024fea, matthewMiller2019fea, thunes2015fea, williamson2023propfea}. Nevertheless, refinements to existing formulations may still be needed to accurately reproduce the extent of bulk and shear deformations revealed in our data.

\subsection{VSI-inferred models exhibited independence from high-order $\overline{I_1}$ and $\overline{I_2}$ terms}
\label{Sect:DiscVSIParams}
The VSI framework identified a pure m-HGO model and a polynomial model comprising only three terms: a neo-Hookean form plus an anisotropic component with second-order dependence on $I_4$. Remarkably, this reduced polynomial model successfully captured key internal features of both intact and torn tendons, as evidenced by their displacement distributions, and performed comparably to the m-HGO model. This highlights the ability of VSI to extract a parsimonious representation of the strain energy density function. A compact, three-term model may offer advantages for preliminary investigations of tendon behavior, particularly by avoiding the challenges associated with calibrating exponential-based formulations like the m-HGO model. Although the m-HGO form also demonstrated independence from higher-order terms in $I_1$ and $I_2$, its exponential structure inherently introduces a nonlinear dependence on, at least, the first invariant.

\subsection{PDE-constrained inference yielded physically plausible parameters}
\label{Sect:DiscPDEConstrainedParams}
Tendons are assumed to be highly incompressible materials, with Poisson’s ratios approaching 0.5. However, the parameters inferred by VSI did not reflect this behavior. In most tendons, the inferred bulk modulus was smaller than the corresponding shear modulus near the imposed boundary, which would imply a negative Poisson’s ratio or unphysical result. In contrast, the PDE-constrained optimizer successfully identified parameters consistent with the incompressible regime. While a key advantage of VSI lies in its ability to identify active deformation mechanisms and suppress irrelevant terms, its inability to reliably infer the bulk modulus in nearly incompressible materials is a known limitation \citep{wang2021vsipde}.

VSI-inferred shear moduli were on the order of $10^{-1}$ MPa, which is either two to three orders of magnitude higher \citep{williamson2023propfea, garcia2024fea} or two orders of magnitude smaller \citep{quental2016fea} than reported values for healthy human rotator cuff tendons calibrated with an HGO model. After refinement, these values decreased by one to two orders of magnitude, aligning more closely with prior experimental findings in human tendons with smaller shear moduli \citep{garcia2024fea, williamson2023propfea}. The refined $k_1$ parameter fell within reported ranges for the rotator cuff, while $k_2$ remained in the lower nonlinearity regime (i.e., values below 0.1), similar to those found in the anterior bursal region or at the tendon insertion of the human supraspinatus \citep{garcia2024fea, williamson2023propfea}. Given that this is the first study to characterize full volume strain maps in an animal model of the rotator cuff, the lower $k_2$ values may reflect the relatively large contribution of tissue near the insertion at the global scale. However, in the absence of published material parameters for the ovine rotator cuff, our study provides the first reference values for this tissue.

The reduced polynomial model lacks explicit nonlinear and fiber dispersion terms present in the m-HGO model. However, the magnitude of $\theta_7$, which corresponds to a second-order dependence on the fiber field, was comparable to that of $k_1$. This suggests that $\theta_7$ plays an analogous role in capturing fiber contributions. More importantly, this finding implies that modeling the fiber direction field is crucial to achieve physiologically realistic constitutive responses.

\subsection{PDE-constrained optimization enhanced internal strain predictions in intact and torn tendons}
\label{Sect:DiscPDEParams}
The sliding behavior between detached and attached tissue bundles in the torn condition was accurately reproduced following PDE-constrained optimization. Although displacement maps appeared visually similar across the surface for all models, internal shear strain distributions derived from these displacements revealed notable differences. Specifically, the NH model failed to reproduce the distinct shear strain bands observed in the internal regions, particularly near the interface between cut and uncut tissue. In the intact condition, the NH model exhibited a uniform pattern of negligible shear strain, whereas both the m-HGO and Pol models captured clear strain gradients. Similarly, across both inference and validation datasets, strain magnitudes and squared $L_2$ norm values were generally of comparable order across models; however, the largest percentage reductions were achieved by the m-HGO and Pol models. These findings underscore the importance of anisotropic contributions to tendon mechanics, even under low-strain loading conditions, specially for more complex models that include contact between bone and tendon, which have represented tendons as isotropic materials \citep{quental2020suture}.

\subsection{Invariant-specific error maps revealed limitations in volumetric and fiber-related terms}
\label{Sect:DiscErrorMaps}
Although the widely used m-HGO model effectively captures the high shear strain trends observed in our tendon data, localized regions of elevated error remain. Importantly, the largest source of invariant error was associated with $I_3$, suggesting that the strain energy density function could be improved by refining the volumetric contribution. This is a direction that has been explored in other studies using full volume inverse modeling techniques \citep{luetkemeyer2021constmod}. Additionally, $I_4$ error maps revealed concentrated error near the footprint detachment and the tendon gripper, indicating that inaccuracies in the fiber direction field may contribute to these discrepancies. Enhancing the representation of fiber distribution could therefore improve model performance in these regions.

\subsection{Limitations and Future Work}
\label{Sect:DiscLimitFuture}

This study has several limitations, which should be considered when utilizing the inferred parameters for future computational simulations. 

First, in the PDE-constrained optimization, parameter values for the shear modulus, $k_1$, $k_2$, $\kappa$, and $\theta_7$ were often bounded from below to ensure convergence. These lower bounds were frequently reached during optimization, suggesting that the identified parameter sets may represent a local minimum. It is possible that further improvement could be achieved by addressing volumetric locking effects associated with high incompressibility. One potential solution involves using a mixed finite element formulation with both displacement and pressure fields as unknowns. However, because the anisotropic term in our models is not fully deviatoric, adapting this approach would require modifying the strain energy density function. Such modifications could impair the fidelity of fiber-related responses, which is why we chose not to pursue this path in the current study.

Second, the fiber direction field was estimated through interpolation between the articular and bursal edges of the tendon on two-dimensional anatomical image slices. This approach assumed that fibers aligned relative to the outer tendon shell, conforming to the curvature of the humeral head, which matched our high-resolution anatomical observations and has been similarly used in previous studies \citep{luetkemeyer2021constmod, quental2016fea}. However, higher-resolution imaging modalities, such as nano-computed tomography, could provide more precise fiber direction fields and thereby improve constitutive predictions.

Third, we assumed that material properties remained unchanged after detaching a portion of the footprint. This assumption requires further validation, particularly given that torn tendons showed better agreement with model predictions than intact tendons, which is  possibly because of changes in the dominant deformation mechanisms contributing to the VSI and PDE-constrained optimization loss terms. Moreover, shear strain was particularly prominent in torn tendons, while volumetric strain patterns suggested that the tissue exhibits a degree of compressibility. However, the inferred parameters indicated a largely incompressible material behavior. This apparent contradiction may reflect large deformations near the tendon gripper and between detached and attached tissue regions, rather than true volumetric changes, and a different constitutive law in the torn tendon. Future work in this direction will involve inferring constitutive parameters from intact datasets and incorporating them into a phase-field fracture modeling framework, using torn tendons for validation. This approach will help account for fiber remodeling near tear edges and volumetric changes in the cut/uncut tissue interface, and capture injury-induced changes in material parameters more accurately by introducing parameter dependence on the spatially varying phase field for damage.

Additionally, the use of a coarse mesh was driven by the resolution of our MRI data and the noise amplification observed in finer meshes. To mitigate this, we employed second-order tetrahedral elements. Still, higher-resolution experimental data with improved signal-to-noise ratios could enhance model fidelity by capturing internal features more accurately.

Parameter identifiability could also be improved by expanding the validation dataset. Our current setup included only a single quasi-uniaxial experiment. Incorporating additional loading configurations, such as varying abduction or rotation angles, would likely activate additional deformation mechanisms and enhance the robustness of the inference process.

While our study demonstrates that a single material model can effectively capture complex features across the entire tendon, future improvements in performance may be achieved through region-specific modeling approaches, as proposed in prior studies \citep{matthewMiller2019fea, garcia2024fea, szczesny2012mechprop, thunes2015fea}. Such approaches would also facilitate a more detailed understanding of the individual contributions of distinct anatomical regions, such as the enthesis, bursal, and articular sides, to overall tendon deformation. However, adding more regions would also introduce a larger risk for overfitting and unneeded complexity, and should be carefully considered to obtain a true parsimonious representation. Perhaps a better approach would be to consider time-dependent behavior and biological factors into the modeling approach, which was not included in our hyperelastic models.

Finally, our future work aims to incorporate machine learning approaches to develop model-agnostic predictions. In particular, we plan to explore whether strains can be encoded into anatomical shapes using geometric learning methods (e.g. mesh-based networks) that take the 3D geometry under known loading conditions as input and learn a mapping to spatial strain patterns. A neural network operating on the mesh can embed local geometric features like thickness or curvature into a latent space, from which strain hotspots are decoded. Such an approach would potentially enable the identification of strain concentration regions directly from high-resolution images, bypassing the need for explicit strain energy density constitutive forms.

Despite these limitations, our study is the first to use full-volume experimental measurements to calibrate tendon-specific constitutive models of the rotator cuff. We demonstrate that internal deformation patterns can be reproduced using models with a minimal yet essential anisotropic contribution to the strain energy density function, highlighting the complex and heterogeneous mechanical response of the rotator cuff. Our non-destructive characterization approach shows that internal tissue behavior can be captured without tendon excision and suggests that the rotator cuff may be effectively modeled as a continuum with homogeneous material properties in a simplified scheme. This framework lays the foundation for future data-driven tendon modeling tools that integrate patient-specific geometry and loading to guide clinical management.

\bibliographystyle{elsarticle-harv} 
\bibliography{references}


\appendix
\section{Supplemental materials}
\label{Sect:SupFigs}

\begin{figure}[h!]
    \centering
    \includegraphics[width=\linewidth]{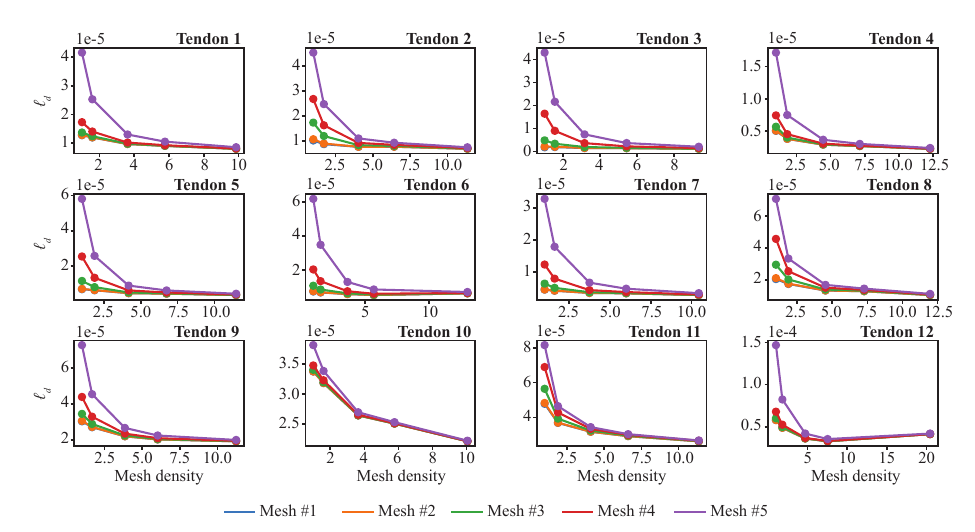}
    \caption{Displacement loss as a function of mesh density, where mesh density is defined as the ratio between the number of elements in a given mesh divided by the number of elements in the coarsest mesh. Each curve shows the loss computed using parameters inferred via VSI for the corresponding labeled mesh.}
    \label{fig:DisplacementLoss}
\end{figure}

\begin{figure}[h!]
    \centering
    \includegraphics[width=\linewidth]{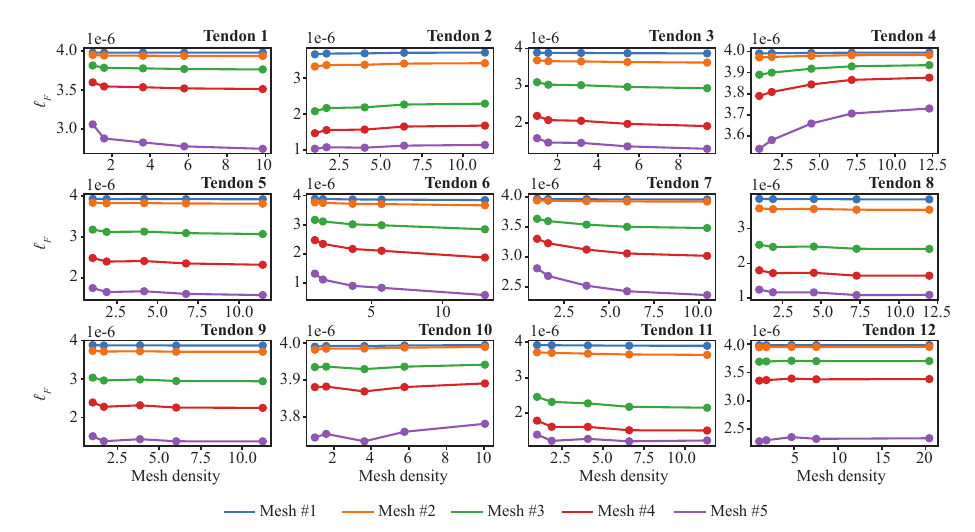}
    \caption{Load loss as a function of mesh density, where mesh density is defined as the ratio between the number of elements in a given mesh divided by the number of elements in the coarsest mesh. Each curve shows the loss computed using parameters inferred via VSI for the corresponding labeled mesh.}
    \label{fig:LoadLoss}
\end{figure}


\begin{table}[h!]
    \centering
    {
    \renewcommand{\arraystretch}{1.2} 
    \setlength{\tabcolsep}{2pt} 
    \begin{tabular}{c c c c c c c}
        \hline
        \multicolumn{7}{c}{Candidate model, $W_{\text{m-HGO}}$} \\
        \hline
        \multirow{2}{*}{Tendon} & \multirow{2}{*}{Mesh type} & \multicolumn{5}{c}{Inferred parameters} \\
        \cline{3-7}
        & & $K$ (MPa) & $\mu$ (MPa) & $k_1$ (MPa)& $k_2$ (x$10^{-1}$) & $\kappa$ \\
        \hline
        \multirow{2}{*}{1} & Fine mesh, linear element & 0.20 & 0.58 & 0.80 & 6.79 & 0 \\
        & Coarse mesh, quadratic element & 0.20 & 1.08 & 1.40 & 6.84 & 0 \\
        \hdashline
        \multirow{2}{*}{2} & Fine mesh, linear element & 0.20 & 0.16 & 1.13 & 2.69 & 0 \\
        & Coarse mesh, quadratic element & 0.20 & 0.38 & 1.42 & 2.65 & 0 \\
        \hdashline
        \multirow{2}{*}{3} & Fine mesh, linear element & 1.98 & 0.82 & 1.00 & 2.26 & 0 \\
        & Coarse mesh, quadratic element & 1.70 & 1.36 & 3.62 & 1.23 & 0 \\
        \hline
    \end{tabular}}
    \caption{Comparison of parameters inferred for the m-HGO model, using the VSI framework and different types of meshes. Three tendons are shown.}
    \label{tab:ComparisonCoarseFineMesh}
\end{table}







\end{document}